\documentclass[english,prd,nofootinbib,preprintnumbers,twocolumn,showpacs]{revtex4}
\usepackage[english]{babel}
\usepackage[latin1]{inputenc}
\usepackage{amsmath}
\usepackage{amsfonts}
\usepackage{appendix}
\usepackage{amssymb}
\usepackage{epsfig}
\usepackage{graphics,psfrag,rotating}
\usepackage{graphicx}
\usepackage{dcolumn}
\usepackage{bm}
\usepackage{epstopdf}
\usepackage{color}
\usepackage[usenames,dvipsnames,svgnames]{xcolor}
\usepackage[colorlinks=true,
            linkcolor=red,
            urlcolor=gray,
            citecolor=blue]{hyperref}
\usepackage{hyperref}
\usepackage[T1]{fontenc}
\usepackage{multirow}
\usepackage{float}
\usepackage{hyperref}

\usepackage{subfigure}

\usepackage{enumitem}

\def\3nab{\tilde{\nabla}}

\def\be {\begin{equation}}
\def\ee {\end{equation}}
\def\ba {\begin{align}}
\def\ea {\end{align}}

\def\bc {\begin{center}}
\def\ec {\end{center}}
\def\case#1/#2{\frac{#1}{#2}}

\newcommand{\bver}{\begin{verbatim}}
\newcommand{\ever}{\end{verbatim}}

\newcommand{\bea}{\begin{eqnarray}}
\newcommand{\eea}{\end{eqnarray}}
\newcommand{\beaa}{\begin{eqnarray*}}
\newcommand{\eeaa}{\end{eqnarray*}}

\def\case#1/#2{\textstyle\frac{#1}{#2}}

\begin{document}
\title{On neutron stars in $f(R)$ theories: small radii, large masses \\
and large energy emitted in a merger}

\author{
Miguel Aparicio Resco$^{a}~\footnote{migueapa [at] ucm.es}$, 
\'Alvaro de la Cruz-Dombriz$^{b}~\footnote{alvaro.delacruzdombriz [at] uct.ac.za}$, 
Felipe J. Llanes-Estrada$^{a}~\footnote{fllanes [at] fis.ucm.es}$ and
V\'ictor Zapatero Castrillo$^{a}~\footnote{victorza [at] ucm.es}$
}
\affiliation{
$^{a}$ Departamento de F\'{\i}sica Te\'orica I, Ciudad Universitaria, Universidad Complutense de Madrid, E-28040 Madrid, Spain.\\
$^{b}$ Astrophysics, Cosmology and Gravity Centre (ACGC), Department of Mathematics and Applied Mathematics, University of Cape Town, Rondebosch 7701, Cape Town, South Africa.
}



\begin{abstract} 
In the context of $f(R)$ gravity theories, we show that the apparent mass of a neutron star as seen from an observer at infinity 
is numerically calculable but requires careful matching, 
first at the star's edge, between interior and exterior solutions, none of them being totally Schwarzschild-like but presenting instead small oscillations of the curvature scalar $R$;
and second at large radii, where  the Newtonian potential is used to identify the mass of the neutron star.
We find that
for the same equation of state,
this 
mass definition is always larger than its general relativistic counterpart. We exemplify this with quadratic $R^2$ and Hu-Sawicki-like modifications of the standard General Relativity action. Therefore, the finding of two-solar mass neutron stars basically imposes no constraint on stable $f(R)$ theories. However, star radii 
are in general smaller than in General Relativity, which can give an observational handle on such classes of models at the astrophysical level.
Both larger masses and smaller matter radii are due to much of the apparent effective energy 
residing in the outer metric for scalar-tensor theories. 
Finally, because the $f(R)$ neutron star masses can be much larger than General Relativity counterparts, the total energy 
available for radiating gravitational waves could be of order several solar masses, and thus a merger of these stars constitutes an interesting wave source.
\end{abstract} 


\maketitle

\section{Introduction}

\label{Section1}

General Relativity (GR) is the currently widely accepted theory of gravity, adopted by both astrophysicists and cosmologists. This theory has shown to be predictive in a wide range of situations, from the highly accurate Solar system tests - including orbital precession, gravitational Doppler effect and light bending, among others - to some  extragalactic features as the indirect detection of gravitational waves emission through binary pulsar systems 
({\it c.f.} the extensive review by Will \cite{Will:2014kxa} about the extraordinary success of GR as well as proposal for future tests). Nevertheless, GR exhibits some limitations too, since within this theoretical frame some important problems remain unsolved. Undeniable examples are the so called {\it dark matter problem} in the context of astrophysics, the accelerated expansion of the Universe, i.e., the  {\it dark energy problem}, and the early inflation, both in the context of Cosmology. These and other problems lead into the search of extended theories of gravity that might succeed at describing the situations in which the GR fails. Such extended theories must be indeed extensions of GR capable of recovering 
the aforementioned classical tests and explain or cure the open issues in Cosmology and Gravitation.
In this context, $f(R)$ gravity theories remain one of the simplest and most successful ways to extend the GR and the one we adopt in this paper ({\it c.f.} \cite{Carroll:1997ar} for extensive reviews). 
The increasing interest on these theories lies in their ability to describe gravitating structures by 
providing the main contribution to the non-baryonic dark matter of the Universe
by means of the extra scalar mode ({\it c.f.}  \cite{Cembranos:2008gj, Capozziello:2012ie} for explicit examples), together with the possibility to unify the cosmic speed-up \cite{delaCruzDombriz:2006fj} and the early-time inflation through, for instance, the Starobinsky inflationary $R^2$-model \cite{Planck-Inflation}, thus leading to a complete picture of the evolution of the Universe \cite{staro, applications_staro} and large-scale structures therein \cite{perturbations}.

The scenario to be thoroughly explored in this paper is the strong gravitational field regime. Namely we shall study the existence and main features of neutron stars in the context of $f(R)$ paradigmatic models. In order to describe these archetypical gravitational configurations realistically, we shall consider up-to-date nuclear equations of state (EoS) 
that effectively account for the interacting neutrons inside the star from a microscopic basis \cite{Gezerlis:2013ipa}.
As widely reported in the literature, once GR is assumed as the underlying theory of gravity, neutron star masses are bounded from above. The theoretical mass limit itself has increased with time from $\simeq 0.6 M_{\odot}$ that  Tolman-Oppenheimer-Volkoff (TOV)~\cite{Oppenheimer:1939ne} found for a free neutron gas EoS, to the 1.5-1.6 solar masses found with traditional potential models of the nuclear force, to $\mathcal{O}(2.2M_{\odot})$ limits found with stiffer chiral interactions~\cite{Dobado:2011gd}. 

Notwithstanding, the discovery  of a neutron star with mass $1.974M_{\odot}$ \cite{Demorest:2010bx} by the Shapiro delay method, confirmed by a second very precise sighting~\cite{Antoniadis:2013pzd} through binary system measurements, as well as various other results in different configurations~\cite{others_masses}, have seemed to rule out already many exotic EoS.
Some attempts trying to reconcile these results with GR and softer EoS have for instance assumed
strong magnetic fields inside the star ({\it c.f.} \cite{Astashenok:2014nua} for a detailed description of this effect and more specific bibliography) or assumed electric charge~\cite{Jing:2015ota}. As a matter of fact, literature on these supermassive pulsars is growing rapidly.

In this paper we shall show that upper bounds akin to the TOV limit in GR do not exist
for vast classes of $f(R)$ theories
since the consideration of different parameter values for a given class of models 
leads to heavier and heavier masses. Also for some specific models
and given EoS, the mass-radius diagrams seem not to be bounded from above. In other words, as
we shall see in the corresponding mass-radius relations, 
higher maximum masses can often be reached by a simple variation of the free parameters values for the classes of $f(R)$ models under study.  Mass-radius relations in the realm of extended theories of gravity have been previously addressed in the literature. 
In this regard, a popular approach has consisted of solving the extended field equations by means of perturbative techniques, aimed indeed by the complexity of the 
higher-order equations as well as the impossibility of getting a decoupled first-order equation for the pressure as in GR \cite{Oppenheimer:1939ne}. Perturbative techniques were used in \cite{Arapoglu:2010rz}, where the $R^2$ model was considered, or~\cite{Astashenok:2013vza}, where the authors explore some particular models as $f(R)=a R^2$, $f(R)=a R^{2}(1+ a R)$ and $f(R)=a R^{2}(1+ b \log(\frac{R}{\mu^2}))$. Obvious limitations of a perturbative approach may be enumerated: among others, the impossibility of comparing the unknown exact solution with the determined perturbative solution, for not mentioning that in strong gravity regimes, gravitational effect beyond GR are expected to be dominant and consequently, the GR limit cannot be simply considered as the leading contribution in the solution. 
A different method is used in \cite{Yazadjiev:2014cza}. In this case, the equivalence between $f(R)$ theories and Brans-Dicke scalar-tensor theory is used to numerically solve the modified field equations with the corresponding potential for the scalar field. Some cosmological stability constraints are assumed and boundary conditions are imposed in the numerical shooting method. Finally, a sort of exact numerical solutions of the modified equations  were found in \cite{Capozziello:2015yza} for gravitational  Lagrangians of the form $f(R)=R+ a R^{2}(1+\gamma R)$ and $f(R)=R^{1+\epsilon}$. Although the latter formulation is more similar to the approach that we shall develop herein, the methods used in the references 
above, still share 
a common denominator:  the definition for the star gravitational mass remains therein 
just inherited from the standard definition for mass in GR (see the Appendix \ref{App:B}). 
Nevertheless, as shall be shown in the bulk of this article, $f(R)$ gravity theories do not admit in general a trivial generalization of the mass definition. 
Even though the definition widely used in the literature can still be thought of as a theoretical label for the neutron star, as it could in the GR case, it can by no simple means be measured through astrophysical observations. As a consequence,  the usefulness of the resulting mass-radius diagrams in the existent literature remains limited. 
In order to tag the neutron stars in $f(R)$ theories with an observable mass, in the following
we shall suggest a measurable gravitational mass and construct the diagrams accordingly.

 A direct consequence of such analysis lies in the possible differences in gravitational-wave signals of stellar mergers respect to their GR counterparts.
%
The existence of gravitational waves in $f(R)$ theories has already been demonstrated \cite{GW_f(R)} and the main peculiarity thereof is the propagation of an additional longitudinal mode, surplus to the transverse ones already present in GR. This corresponds to the propagation of the 
scalar field in the canonically equivalent scalar-tensor modified gravity theories.
Beyond-GR theories, even for
the simplest static and spherically symmetric configura-
tions, distort space-time in novel ways; in the bulk of the paper
we shall see that standing waves of the $R$ scalar form in the
vacuum outside a star. In a merger, interesting gravitational signals,
particularly gravitational waves, will propagate out.
Provided a deficit between
the \emph{quantity of matter} and the \emph{total mass} of the neutron star happens,  the merger of two compact objects in the context of $f(R)$ theories 
may provide a stronger signal than if the calculations were performed in the GR context, where black holes are usually the required merging objects in order for the wave signal to be detected. In other words, we shall show how whenever gravity is significantly modified at neutron star scale, which certainly remains to be seen,
then $f(R)$ theories could naturally accommodate a 3-4 solar mass emission without resorting to black-hole merging as for instance claimed recently by LIGO collaboration \cite{LIGO, LIGO2}.

The paper is organised as follows: 
In Sec.~\ref{Section2} we shall first introduce the  
key equations in the metric formalism of $f(R)$ theories including the dynamical set to be integrated. 
Next in \ref{subsec:solution} 
we shall briefly present the 
analysis which ensures the stability of the exterior solutions. Then, we shall argue the boundary conditions that are inevitable to set out a well-posed problem with physical solutions. 
To conclude that section, in  \ref{sec:eqofstate} 
we shall quickly examine the state of the art for the EoS 
for the matter content (saliently the neutron fluid) in the star, and specify the modern versions of the EoS 
we have examined and subsequently adopted throughout the paper.  
In Sec.~\ref{sec:waves} we shall study the propagation velocities for the metric tensor perturbations  
as functions of the $f(R)$ induced corrections. 
Sec.~\ref{sec:mass}
then introduces the suitable gravitational mass by means of a straightforward comparison and extension of the GR case.
Immediately afterwards, in Sec. \ref{sec:numerics} 
 we shall carefully describe the numerical procedure to identify physical masses that, for the sake of simplicity, we shall illustrate 
 within a paradigmatic example, $f(R)=a R^{2}$. To speed the presentation we start with a small value of $a$ that, not being yet too far from GR allows to have 
one less degree of freedom in the initial conditions
at the price of satisfying only approximately one of the boundary conditions. Then we shall lift the approximation and solve the system in full generality. 
Secs.~\ref{SectionRsq} and \ref{SectionHS} 
are  then devoted to presenting the mass-radius diagrams obtained for several $f(R)$ classes of models using realistic neutron-star EoS.  
Immediately after in Sec.~\ref{sec:GW}
we present a brief description of the potential implications that the obtained results and mass labelling may have in the gravitational-wave detection interpretation
in the $f(R)$ theories realm.
 Outlook and discussion of the results are provided in Sec.~\ref{sec:Conclusions}. 
Finally, the derivation of the system of dynamical equations, as well as some more comments on the assignment of mass in $f(R)$ theories, are relegated to the Appendices~\ref{App:A} and~\ref{App:B} respectively.

\section{Analysis of static and spherically symmetric space-times in $f(R)$ theories}
\label{Section2}
We promote the Einstein-Hilbert Lagrangian for a space-time with scalar curvature $R$ to a generic function $f(R)$, 
so the gravitational action becomes
\begin{equation}\label{1}
{S}=\frac{1}{2\kappa}\int {\rm d}^4 x \sqrt{-g}\,[R+f(R)],
\end{equation}
where $\kappa=\frac{8 \pi G}{c^{4}}$. 
We obtain  the corresponding $f(R)$ Einstein field equations by varying with respect to the metric\footnote{
Note that our definition for the Riemann tensor is
$R^{\sigma}_{\mu \nu \kappa}=\partial_{\kappa} \Gamma ^{\sigma}_{\mu \nu}-\partial_{\nu} \Gamma ^{\sigma}_{\mu \kappa}+\Gamma ^{\sigma}_{\kappa \lambda}\Gamma ^{\lambda}_{\mu \nu}-\Gamma ^{\sigma}_{\nu \lambda}\Gamma ^{\lambda}_{\mu \kappa}$.
},
\begin{eqnarray}\label{2}
R_{\mu\nu}-\frac{1}{2}\,R\,g_{\mu\nu}=\frac{1}{1+f_{R}}\,[-\kappa\,T_{\mu\nu}-\nabla_{\mu}\nabla_{\nu}f_{R}\nonumber\\
+\,g_{\mu\nu}\,\nabla^{\alpha}\nabla_{\alpha}f_{R}+\frac{1}{2}(f(R)-R\,f_{R})\,g_{\mu\nu}],
\end{eqnarray}
where $f_{R}\equiv{\rm d}f(R)/{\rm d}R$ (likewise $f_{2R}$, $f_{3R}$ and $f_{4R}$ in the following will denote higher derivatives) and the energy-momentum tensor is  
\begin{equation}\label{1a}
T_{\mu\nu}=\frac{2}{\sqrt{-g}}\frac{\delta (\mathcal{L} _{matter} \sqrt{-g})}{\delta g^{\mu\nu}},
\end{equation}
The contraction of  Eq.~(\ref{2}) with $g^{\mu\nu}$ and some rearranging provides the expression of the scalar curvature as follows
\begin{eqnarray}\label{3}
R=\frac{\kappa\,T-2\,f(R)-3\,\nabla^{\alpha}\nabla_{\alpha}f_{R}}{1-f_{R}}\ .
\end{eqnarray}
Unlike GR, where $R$ and $T$ are algebraically constrained by $R=\kappa\,T$, 
the $f(R)$ action makes Eq.~(\ref{3}) a dynamical, differential relation 
between the matter sources and the derivatives of the Lagrangian $f$. 

Being interested in static neutron stars, we consider the most general static and spherically symmetric four-dimensional metric tensor
\begin{equation}\label{4}
{\rm d}s^{2}=B(r)\,{\rm d}t^{2}-A(r)\,{\rm d}r^{2}-r^{2}({\rm d}\theta^{2}+\sin^{2}{\theta}\,{\rm d}\phi^{2}),
\end{equation}
and assume that its matter content is describable by a perfect fluid~\footnote{See for example~\cite{Manuel:2014kqa} for work and references on nonperfect fluidity and transport.
} whose energy-momentum tensor in the comoving frame can be written as
\begin{eqnarray}
T_{\mu\nu}=\left(\rho+p\right)u_{\mu}u_{\nu}-p g_{\mu\nu}\,. 
\label{Tmunu}
\end{eqnarray}
Then, substitution of Eqs. (\ref{4}) and (\ref{Tmunu}) in the field equations~(\ref{2}) as well as the conservation of the energy-momentum $\nabla_{\mu}T^{\mu\nu}=0$ render the following set of independent equations (see Appendix \ref{App:A} for the exhaustive derivation),
\begin{eqnarray} \label{5}
A'&=&\frac{2rA}{3(1+f_{R})}\left[\kappa A(\rho+3p)+\frac{A}{2}R-\frac{3B'}{2rB}+Af(R)\right.\nonumber\\
&&\left.-\,f_{R}\left(\frac{A}{2}R+\frac{3B'}{2rB}\right)-\left(\frac{3}{r}+\frac{3B'}{2B}\right)f_{2R}\,R'\right]\,,
\\
\nonumber\\
\nonumber\\
%
\label{7}
B''&=&\frac{B'}{2}\left(\frac{A'}{A}+\frac{B'}{B}\right)+\frac{2A'B}{rA}+\frac{2B}{(1+f_{R})}\left[-\kappa Ap\right.\nonumber\\
&&\left.-\frac{A}{2}R+\left(\frac{B'}{2B}+\frac{2}{r}\right)f_{2R}\,R'-\frac{A}{2}f(R)\right],
\\
\nonumber\\
\nonumber\\
%
\label{6}
R''&=&R'\left(\frac{A'}{2A}-\frac{B'}{2B}-\frac{2}{r}\right)-\frac{A}{3f_{2R}}\left[\kappa(\rho-3p)\right.\nonumber\\
&&\left.-(1-f_{R})R-2f(R)\right]-\frac{f_{3R}}{f_{2R}}\,R'^{2},
\\
\nonumber\\
\nonumber\\
%
\label{8}
p'&=&-\frac{\rho+p}{2}\,\frac{B'}{B}.
\end{eqnarray}
where the prime denotes a derivative with respect the radial coordinate $r$. The functions to be determined by this system of equations are $A(r)$, $B(r)$, $R(r)$ and $p(r)$, the latter being in principle not zero inside the star, as it vanishes by definition in its exterior. In order to close the system, a fluid EoS $\rho=\rho(p)$ is required. 
In this regard our choices will be discussed shortly in Sec.~\ref{sec:eqofstate}.

The system of Eqs.~(\ref{5})-(\ref{8}), together
with the trivial definitions $B''={\rm d}B'/{\rm d}r$ and $R''={\rm d}R'/{\rm d}r$
become a set of six first-order, non-linear coupled ordinary differential equations, 
that we have solved numerically using a fourth-order Runge-Kutta algorithm.
We have specified the initial conditions at the centre of the star and integrate outwards until the pressure vanishes. At this point the star's edge has been reached and from there we set $p=0$, so that in the exterior of the star the system contains one differential equation less, and the EoS is not needed any longer.
Thus we 
continue integrating outwards to a large enough distance, as explained below in Sec.~\ref{sec:mass}.

\subsection{Perturbative analysis and Boundary conditions} 
\label{subsec:solution}

Since GR appears to be a sound, at least approximate framework to deal with compact stars, we first perform a perturbative analysis in the scalar curvature around the GR solution $R_0$ such that $R=R_{0}+R_{1}$ with 
\begin{equation}\label{expandirR}
R_{1}\ll R_{0}=\kappa T\ . 
\end{equation}
This pertubation must remain bounded and avoid rapid growth to guarantee solution stability.
We substitute this perturbation on Eq.~(\ref{6}) and keep first-order terms in $R_{1}$. 
As $R_0$ on the right-hand side is known from Eq.~(\ref{expandirR}), the outcome is a dynamical equation for $R_1$
as follows
\begin{eqnarray}\label{16}
R_{1}''&=& - R_{0}'' + R_{0}'\left(\frac{A'}{2A}-\frac{B'}{2B}-\frac{2}{r}\right)-\frac{f_{3R}(R_{0})}{f_{2R}(R_{0})}\,R_{0}'^{2}\nonumber\\
&&-\frac{A}{3f_{2R}(R_{0})}\left(R_{0}f_{R}(R_{0})-2f(R_{0})\right)
+\,\delta\,R_{1}'+\gamma\,R_{1},\nonumber\\
\end{eqnarray}
where the coefficients $\delta$ and $\gamma$ are defined as
\begin{eqnarray}\label{17}
\delta=\frac{A'}{2A}-\frac{B'}{2B}-\frac{2}{r}-\frac{2\,R_{0}f_{3R}(R_{0})}{f_{2R}(R_{0})}\,,
\end{eqnarray}
\begin{eqnarray}\label{18}
\gamma&=&\frac{A}{3f_{2R}(R_{0})}\left[1+R_{0} f_{2R}(R_{0})+f_{R}(R_{0})\right.\nonumber\\
&&\left.-\,\frac{f_{3R}(R_{0})}{f_{2R}(R_{0})}\left(2f(R_{0})+R_{0}f_{R}(R_{0})\right)\right]\nonumber\\
&&+\,\frac{R_{0}'^{2}}{f_{2R}(R_{0})}\left(f_{4R}(R_{0})-\frac{f_{3R}^{2}(R_{0})}{f_{2R}(R_{0})}\right).
\end{eqnarray}
Particularising now to the exterior solution, i.e., $R_{0}=R_{0}'=0$, the coefficient $\gamma$ becomes
\begin{eqnarray}\label{simplifiedgamma}
\gamma=\frac{A}{3f_{2R}(0)}\left(1+f_{R}(0)-\frac{2f_{3R}(0)\,f(0)}{f_{2R}(0)}\right)\,.
\end{eqnarray}
In order to guarantee that the perturbation does not grow exponentially, the condition $\gamma<0$ needs to be satisfied\footnote{We must reject $\gamma>0$ since in that case the homogeneous part of Eq.~(\ref{16}) $R''_1 = \gamma R_1 + \dots$ would admit a general solution made of a linear combination of both one exponentially decreasing and one exponentially increasing solutions, the latter forcing $R$ to be very far from $0$ in the exterior of the star, and consequently preventing any possible agreement with GR.}
 so that the tension term proportional to $R_1$  in 
 Eq. (\ref{16}) 
 yields the perturbation in $R_1$ to be an oscillatory function of $r$.
This will be important in what follows, and can be seen for example in Fig.~\ref{Figure_5a}.

Next we notice that the parameter $\delta$ is negative in the exterior of the star. Indeed,
according to Eq.~(\ref{17}) since the exterior solution in GR is Schwarzschild-like, the induced perturbation
lies near the well-known solution $B(r)=A^{-1}(r)=1-\frac{2M}{r}$, (see the lower panel of Fig.~\ref{Figure_5a} for an explicit numerical solution), and consequently both $A$, $B$ and $B'(r)$ are positive, $A'(r)$ is negative and $R_0$ vanishes. Substituting all this information in Eq.~(\ref{17}) results in $\delta<0$. 

Thus we conclude that the term proportional to $R_{1}'$ in 
Eq.~(\ref{16}) is in fact a damping term. In conclusion, provided  $\gamma<0$,
a small perturbation of the Schwarzschild solution will perform damped oscillations\footnote{For the case $f(R)=a\,R^{2}$, it is straightforward to show that $\gamma<0$ implies $a<0$.
}. 
Moreover, since $f(R)$ models with $f(0)\neq0$ were shown not to
host an exterior solution which can be matched to a  Schwarzschild
space-time at very large distances~\cite{Dombriz_BH_2009}, we are entitled to focus on scenarios with $f(0)=0$, so that Eq.
~(\ref{simplifiedgamma}) becomes
\begin{eqnarray}\label{20}
\gamma=\frac{A}{3f_{2R}(0)}\left(1+f_{R}(0)\right)<0.
\end{eqnarray}
This stability condition constitutes an {\it a priori} theoretical requirement. In addition to it, 
for each model $f(R)$ under consideration,  as the examples considered below in Secs.~\ref{SectionRsq} and~\ref{SectionHS},  further conditions can ensue from
other astrophysical and cosmological conditions ensuring the viability of the model~\cite{Pogosian_et_al}.

Let's at this stage turn to the physical boundary conditions. 
The numerical integration will be interrupted for matching at radius $r_{\odot}$ for which $p(r_{\odot})=0$ which will be considered as the star radius. The integration for larger radii will proceed in a vacuum  scenario, in other words,  $\rho=p=0$.  Analogously to the standard GR procedure, Minkowski space-time will be asymptotically required, \emph{i.e.},  
$\lim_{r\to \infty}B=\lim_{r\to \infty}A=1$ and $\lim_{r\to \infty}R=\lim_{r\to \infty}R'=0$. 

From the perturbative analysis shown above, the scalar curvature $R$ is expected to show damped oscillations in the exterior, and provided $\gamma<0$, both $R(r)$ and $R'(r)$ will naturally approach zero at large distances. Therefore, there are two remaining {\it boundary} conditions to be satisfied, namely  $\lim_{r\to \infty}B=\lim_{r\to \infty}A=1$, with two free initial conditions, namely the values $B(0)$ and $A(0)$ at the centre of the star.  Further details about these two initial conditions are provided in Appendix~\ref{App:A}. 
 Thus, the solution of the system formed by  Eqs.~(\ref{5})-(\ref{8}) 
is a well-posed initial value problem whenever the solutions $A(r)$ and $B(r)$ 
approach constant values for asymptotically large radii. 

Obviously, if $R_1\sim R_0$ as happens inside and perhaps near the star, the perturbative discussion provides only an illustrative picture and one needs to resort to a full computational analysis to obtain valid solutions. This has been addressed in earlier literature, 
although the treatment needed further improvement because the star mass definition remains cumbersome in $f(R)$ theories. We fully address this caveat in Sec.~\ref{sec:mass}  below and then turn to the actual numerical computations in Secs.~\ref{sec:numerics}, \ref{SectionRsq} and \ref{SectionHS}. 
Since the closeness of the differential system requires a parametrization of the EoS, before targeting the full numerical resolution 
let us say a few words below about realistic EoS.

\subsection{Equation of state of neutron matter}\label{sec:eqofstate}
In our dynamical system of equations 
Eqs.~(\ref{5})-(\ref{8}), we need to specify the EoS of the nuclear matter involved. 
Neutron stars reach densities well above the nuclear saturation density $\rho_0$ which are not directly accessible to nuclear laboratory experiments. Therefore a theory extrapolation is 
necessary and needs to be constrained as much as possible.

Traditionally, the EoS was calculated from nuclear potentials based on models of the nuclear force, such as~\cite{Akmal:1998cf} and~\cite{Heiselberg:1999mq}. In the last decade an effort is underway to put nuclear computations in a better theoretical footing based on Chromodynamics via Effective Field Theories (EFT) thereof. The first step in this program is to provide adequate both nucleon-nucleon and three-nucleon forces and is relatively well understood. Once in possession of them one needs to compute the EoS. One option is to use many body methods, such as Green's functions Monte Carlo~\cite{Gezerlis:2013ipa}. Another is to resort to dispersion relations in neutron matter~\cite{Lacour:2009ej}. In the following we have employed EoS obtained in both ways.

For developing our numerical method in Sec.~\ref{sec:numerics} we have employed 
the EoS of Manuel and Tolos~\cite{Manuel:2011ed} which is based on the more traditional
ones such as Akmal, Pandharipande and Ravenhall~\cite{Akmal:1998cf} which is a very much employed benchmark, and its reparametrization by Heiselberg and Hjorth-Jensen~\cite{Heiselberg:1999mq}. In fact, small effects such as neutron pairing near the surface are neglected in this EoS.

\begin{figure}
  	\includegraphics[width=0.49750\textwidth]{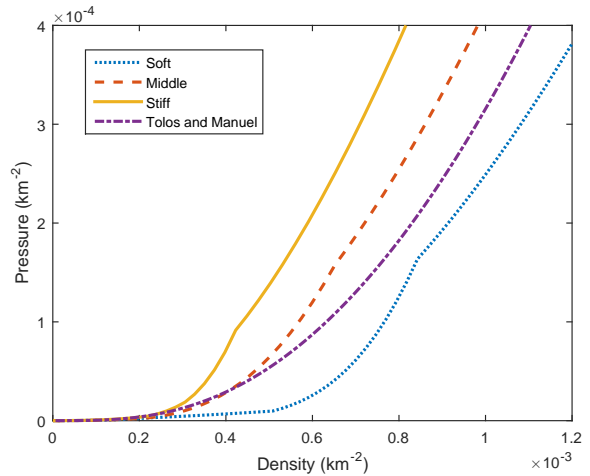}
		\caption{
 Equations of state of~\cite{Manuel:2011ed} (used in Sec.~\ref{sec:numerics}) and~\cite{Hebeler:2013nza} (used in Secs.~\ref{SectionRsq} and~\ref{SectionHS}). Both are neutron-matter EoS, although whereas the first is based on traditional potential models and used only to calibrate our numerical method with a well explored one, the second has also a foundation on EFT methods~\cite{Gezerlis:2013ipa} with better defined errors and consequently we have used it to explore $f(R)$ models in detail.}  
  \label{Figure_5}
\end{figure}

However, in order to explore the $f(R)$ models in more depth, we wish to inch closer to more
 modern and standard-model based chiral EoS. We have found that a good compromise between rigour and practicality, providing a likely uncertainty band for the EoS, is the work of 
~\cite{Hebeler:2013nza}, so we have thus adopted the EoS therein, labelled
``stiff'', ``middle'' and ``soft'' (in reference to the behavior of the pressure as more mass is compacted, i.e., the pickup rate of the sound speed squared $c_s^2 ={\rm d}P/{\rm d}\rho$). For each of
 them we shall systematically swipe the quadratic 
and Hu-Sawicki $f(R)$ models in Secs.~\ref{SectionRsq} and~\ref{SectionHS}. 
These three EoS 
are represented in Fig.~\ref{Figure_5}.

\begin{figure} [H] 
\includegraphics[width=0.4\textwidth]{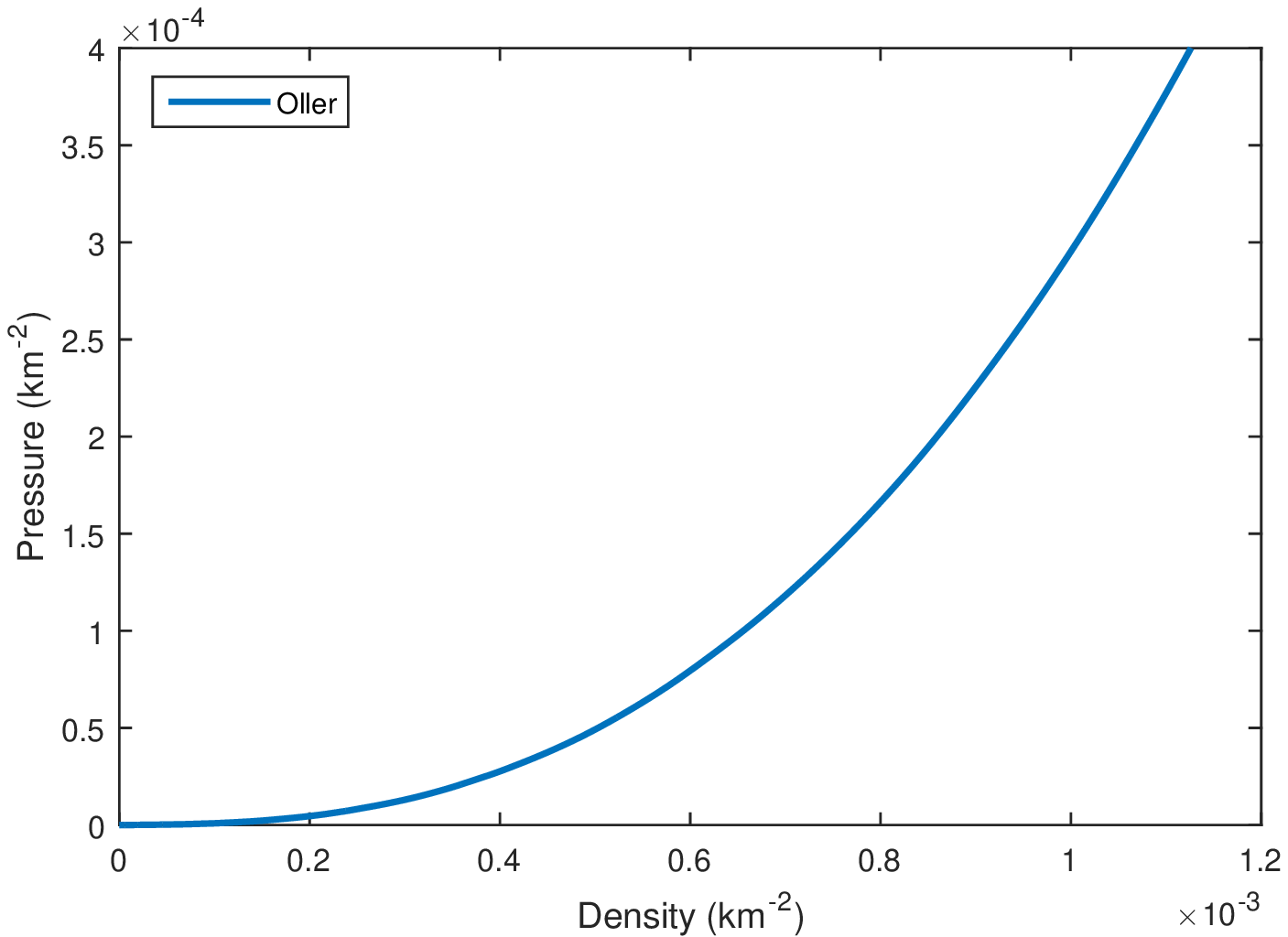}
\includegraphics[width=0.4\textwidth]{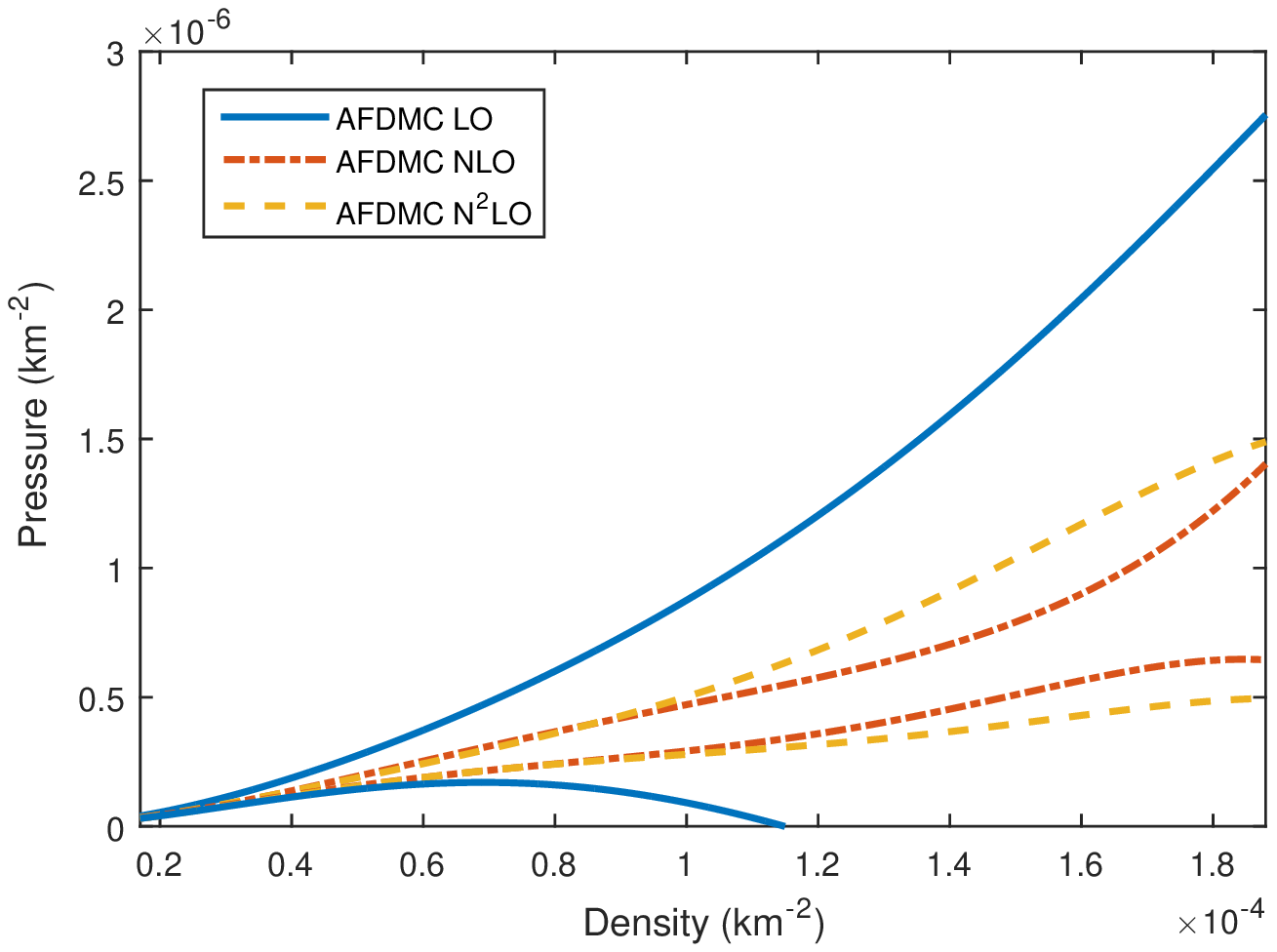}
\caption{
\label{fig:otrasEcsEstado} 
Other modern equations of state by Lacour, Oller and Meissner~\cite{Lacour:2009ej} (top) and by Gezerlis {\it et al.} \cite{Gezerlis:2013ipa} (bottom). 
We do not show results for neutron stars employing these equations but we have also used them and obtained very similar results to the one depicted in the paper.}
\end{figure}

On the other hand, for the sake of completeness we have also performed similar computations (that will not be reported in detail)
with purely chiral EoS based on the works of Lacour, Meissner and Oller~\cite{Lacour:2009ej} 
as applied in~\cite{Dobado:2011gd}, and of Gezerlis {\it et al.}~\cite{Gezerlis:2013ipa}.
While both are based on a similar approach to the nuclear force starting from chiral, QCD-based forces, the first one obtains the EoS therefrom with the help of a dispersive analysis valid at NLO in the interaction, whereas the second employs a Greens function Monte Carlo method and provides LO, NLO, and up to NNLO curves. These additional EoS are represented in Fig.~\ref{fig:otrasEcsEstado}.
We have found no qualitative changes in the behavior of $f(R)$ stars and will refrain from reporting additional plots to maintain a reasonable presentation length. 

As a last comment, we note that we have focused our attention mostly on the purely neutronic EoS, not valid below about half of the nuclear saturation density $\rho_0/2$. 
For this reason,  we have also added for some runs a nuclear crust employing the very detailed work of~\cite{Sharma:2015bna} at lower densities. The contribution of this crust to the total mass of the star is very small and we can safely neglect it, so we have not employed it extensively.
This very low density EoS is in broad, while not very detailed, agreement with that of other authors, such as~\cite{Hebeler:2013nza}. Likewise, if the chiral EoS becomes too stiff so that $c_s>c$, which only happens for extremely high pressures, we set by hand $c_s=c$ which sets the EoS to the stiff-most possible one as well as guarantees causality.


\section{Propagation of metric tensor perturbations in $f(R)$ theories}
\label{sec:waves}
%
As mentioned in the Introduction, the present paper focuses on time-independent (static) solutions. Nonetheless in this Section we shall study the time evolution of perturbations, so the dynamics of stationary solutions, when subject to small perturbations, will be better understood. Thus, we perform a perturbative analysis where the metric tensor is expressed as
\begin{eqnarray}
g_{\mu \nu}=\eta_{\mu \nu}+\epsilon h_{\mu \nu},
\label{21o}
\end{eqnarray}
where $\eta_{\mu \nu}$ holds for the Minkowskian metric, $\epsilon$ is a dimensionless and small constant 
and $h_{\mu \nu}$ encapsulates the form of the perturbation. For this metric (\ref{21o}), the Ricci scalar becomes
\begin{eqnarray}\label{22o}
R \propto \epsilon \, \eta^{\mu \nu} \partial_{\mu} \partial_{\nu} h,
\end{eqnarray}
where  $h=\eta^{\mu \nu} h_{\mu \nu}$. 
Thus, when conveniently rearranged, the linearisation of the Ricci scalar equation
 (\ref{3}) in vacuum ($T=0$) renders
\begin{eqnarray}\label{23o}
(\eta^{\mu \nu} \partial_{\mu} \partial_{\nu} +m_{R}^{2}) \, R=0,
\end{eqnarray}
where the constant quantity $m_{R}^{2}$ is defined as 
\begin{eqnarray}\label{24o}
m_{R}^{2}=\frac{1+f_{R}(0)}{3f_{2R}(0)}.
\end{eqnarray}
Since $m_{R}^{2} \propto \gamma$, with $\gamma$ as given in (\ref{20}), we conclude that ${\rm sign}(m_{R}^{2})={\rm sign}(\gamma)$. 

The propagation of plane-wave perturbations
\begin{eqnarray}\label{25o}
h \propto e^{-i\,(\omega t - \vec{k} \cdot \vec{r})},
\end{eqnarray}
proceeds, upon substitution of  Eq. (\ref{22o}) in (\ref{23o}),  with the dispersion relation 
\begin{eqnarray}\label{26o}
({\bf{k}}^{2}-\omega^{2})^{2}+m_{R}^{2} \, ({\bf{k}}^{2}-\omega^{2})=0,
\end{eqnarray}
which admits two solutions $\omega_{1,2}$, namely
\begin{eqnarray}\label{27o}
\omega_{1}=\bf{k},
\end{eqnarray}
\begin{eqnarray}\label{28o}
\omega_{2}=\sqrt{m_{R}^{2}+\bf{k}^{2}},
\end{eqnarray}
whose group velocities,  
$v=\frac{\partial \omega}{\partial \bf{k}}$,
are
\begin{eqnarray}\label{27o}
v_{1}=1,
\end{eqnarray}
\begin{eqnarray}\label{28o}
v_{2}=\left(1+\frac{m_{R}^{2}}{\bf{k}^{2}}\right)^{-1/2}\,,
\end{eqnarray}
The first mode is also present in GR and propagates at the speed of light. However, the second mode only appears in $f(R)$ theories propagating at speed different from light's. For exponential solutions ($\gamma > 0$)
the extra $f(R)$ mode propagates for any $k$ and with subluminal speed. These exponentially damped solutions were originally obtained in \cite{Yazadjiev:2015xsj} following a similar analysis to the one detailed above\footnote{We have thoroughly tested such solutions in order to check the consistency of our method. Full agreement was found. For the sake of simplicity we do not provide them here.}. 
Since any wavelength is in principle allowed, those modes may be eventually constrained by LIGO-type interferometers. 
On the other hand, for oscillatory solutions ($\gamma < 0$, $m_{R}^{2} < 0$), the second mode solely propagates above the cutoff
 ${\bf{k}}^{2} > |m_{R}^{2}|$ with superluminal velocity $v_{2}$. Since the theory is locally Minskowski, no Special Relativity based experiment is expected to be contradicted. Thus,the superluminal propagation of the metric tensor perturbations could conceivably be detected by direct measurements of gravitational waves. For f(R) classes of models sufficiently close to GR (such as the values of $a$ depicted in Fig. \ref{Figure_Mass_Radius_Diagram}), the wavelength's order of magnitude for the relevant ($\lambda=\frac{2 \pi}{\bf{k}}$) for the superluminal mode would be around $2$ km, being smaller for models closer and closer to GR. 
Taking into account that LIGO's detectors are set up for wavelengths of order $3000$ km \cite{LIGO,LIGO2}, the aforementioned
superluminal waves lie indeed beyond its observation window.

 \section{Apparent Schwarzschild mass perceived by a far observer}
 \label{sec:mass}

In order to be able to define measurable astrophysical quantities, the careful definition of the gravitational mass  in $f(R)$ theories deserves some attention.  
In elementary classical Newtonian mechanics, mass can be understood as the parameter indicating the gravitational force intensity between two bodies, and it can be written down (see Appendix~\ref{App:B}) as a volume integral of a conserved density. 
A way to probe the Newtonian mass of a system by an external observer is to study the orbit
of a massive particle with mass $m$, energy $E$ and angular momentum $L$, and from its once-integrated radial equation of motion
\begin{eqnarray}\label{22}
\frac{1}{2}\,m\dot{r}^{2}+\frac{mL^{2}}{2r^{2}}-\frac{M_{\rm Newton}\,m}{r}=E\,,
\,
\end{eqnarray}
the system's mass $M_{\rm Newton}$ can be determined.

In GR, each Schwarzschild exterior solution is tagged by the mass 
parameter $M_{\rm GR}$, and a straightforward analysis of the geodesics followed by probe particles leads to an expression analogous to  (\ref{22}), namely
\begin{eqnarray}\label{23}
\frac{1}{2}\,m\dot{r}^{2}+\frac{mL^{2}}{2r^{2}}\left(1-\frac{2M_{\rm GR}}{r}\right)-\frac{M_{\rm GR}\,m}{r}=E\, 
\,
\end{eqnarray}
where an extra piece in the centrifugal term appears\footnote{ 
The fact that gravitational masses in the Newtonian theory and in General Relativity are different, may provide some hint that further extensions of the theory of gravity would bring further complications in the adequate definition of mass.
}.
An integral expression for $M_{\rm GR}$ is still useful, see Appendix  \ref{App:B}. Nevertheless 
the determination of the mass for a compact object in the GR paradigm can in principle be done by studying the motion of satellites thereof within a Schwarzschild space-time\footnote{Other methods of measuring the mass such as the Shapiro effect yield the same result~\cite{Demorest:2010bx} thanks to the strong equivalence principle.}.
A useful parametrization of the metric coefficients in Eq. (\ref{4}), 
\begin{eqnarray}\label{24}
B(r)\equiv1-\frac{2M(r)}{r}\;\;,\;\;  A(r)\equiv\frac{1+U(r)}{B(r)}\,, 
\end{eqnarray}
where $M(r)$ and $U(r)$ are arbitrary functions of the radial coordinate, allows us to rewrite a geodesic equation extending Eq.~(\ref{23}) 
to a form valid in $f(R)$ gravity theories as follows
\begin{eqnarray}\label{26}
\frac{m\dot{r}^{2}}{2}\,(1+U(r))+\frac{mL^{2}}{2r^{2}}\left(1-\frac{2M(r)}{r}\right)-\frac{M(r)m}{r}=E\,.\nonumber\\
\end{eqnarray}
Different types of measurements, whenever interpreted within the GR paradigm, provide the same answer for the mass, because the $(1+U(r))$ factor is unity within it.
However the presence of the factor $(1+U(r))$ in $f(R)$ theories shows that they might yield disparate answers.
For example, if the circular motion of a satellite at distance $r$ is used to determine $M$, the answer depends upon the distance to the star in a way determined by this additional function
which is independent of the mass and needs to be determined simultaneously by observation. This feature of Eq.~(\ref{26}) contrasts with Eq.~(\ref{23}) sharply. 
Likewise, if we consider a free-falling test particle of mass $m$ in a radial $L=0$ trajectory, it will be subject to a potential  $-M(r)/r(1+U(r))$.

In order to choose the most appropriate definition of mass in $f(R)$ theory, we note that for the time being GR has been able to stand all tests at stellar-system scale
 and that $f(R)$ extra terms corrections must, if at all, be small and confined to the neighborhood of compact bodies. 
Thus, if we probe the star from far, we need to recover the usual  Newtonian potential $-\frac{M}{r}$, from which we can naturally interpret the total neutron star mass
(procedure employed also in GR to properly interpret the $M$ parameter of the Schwarzschild solutions as the actual Newtonian mass seen by a far observer).

Thus, by matching the radial effective potential to the Newtonian one means that we can identify  the gravitational mass \emph{function}  as
\begin{eqnarray}\label{27}
M_{f(R)}^{}(r)\equiv \frac{M(r)}{1+U(r)},  
\end{eqnarray}
the gravitational mass far from the star being 
\begin{eqnarray}\label{27a}
M_{f(R)}^{\infty}= \lim_{r\to\infty} M_{f(R)}(r). 
\end{eqnarray}
 Eq.~(\ref{24}) shows how, if the static solutions in $f(R)$ are attempted to be interpreted as Schwarzschild ones in GR, $M(r)$ turns out to be a function of the distance and not a constant parameter, a potential observable effect distinguishing $f(R)$ theories from the GR paradigm.
In order to avoid the spread of confusion in the existing literature around this point, in the following we have decided to tag the static, spherically symmetric solutions of $f(R)$ theory, in an appropriate and natural way, by means of  the $M$ parameter defined by Eq.~(\ref{27a})\footnote{In the literature, the star mass is generally taken as the same integral, Eq.~(\ref{massinGR}), as in GR. This is not guaranteed in $f(R)$ theories.}. 

 If the above matching is satisfied then we recover the Schwarzschild metric away enough from the star.
Thus, the metric functions defined in  Eq.~(\ref{24})  need to satisfy 
\begin{eqnarray}
\label{28}
\lim_{r\to\infty} U(r)=0\,,
\\ 
\label{Matinfinity}
\lim_{r\to\infty}\frac{M(r)}{r}=0\,.
\end{eqnarray} 
%

 \section{Numerical analysis}\label{sec:numerics}
In this section we present in detail the numerical procedure to obtain the $f(R)$ mass label for a full static and spherically symmetric solution. The procedure below shall first illustrate how to 
guarantee a well-behaved Schwarzschild limit at sufficient distance from the star. Then we shall focus on the $f(R)$ physical mass label corresponding to every configuration.
Herein the theoretical discussion will be kept valid for any $f(R)$ model, although for illustrative purposes the presented figures will be generated for a realisation of the model $f(R)=aR^2$, whose parameter space shall be swiped in more detail in Sec.~\ref{SectionRsq}. For the present discussion we will fix $a=-0.05\,{\rm km}^2$ (with natural units $c=G=1$ and distances measured in km as the characteristic neutron star scale); as discussed in Sec.~\ref{subsec:solution}, with our sign conventions\footnote{The EoS used is the Manuel-Tolos EoS~\cite{Manuel:2011ed}.} the Hamiltonian requires negative $a$ to be bounded from below (vacuum stability).

As discussed at the end of Appendix~\ref{App:A}, there are two free initial conditions at the star's  center  to start the fourth order Runge-Kutta numerical integration: $R(0)$ and $B(0)$, with all others determined from physically reasonable requirements. These initial conditions are determined by
a shooting method in order to comply with the boundary conditions~(\ref{28}) and~(\ref{Matinfinity}) at infinite distance. In the interest of computer speed we will only vary automatically one of them, $B(0)$, 
to satisfy the boundary condition in Eq.~(\ref{Matinfinity}).
The initial condition for $R(0)$ will be manually given just a few values to satisfy Eq.~(\ref{28}) to a tolerance of 12 \%.
In this section it is sufficient to fix {\it a priori}  $R(0)=\kappa\,T(0)$, i.e., the GR 
value\footnote{ There is no intrinsic difficulty in fixing $R(0)$ to another value. But a different choice inspired in the effects of the extra $f(R)$  terms increases the computer time required and is postponed to the next sections. Nonetheless, well established successes in describing gravitational radiation~\cite{Weisberg:2010zz} in binary systems lead us to think that $f(R)$ effects in the neighobourhood of neutron stars will be subdominant in comparison to GR effects.}. Our theoretical framework is generally valid for arbitrary $f(R)$ theories, but in the specific numerical example of this section  we shall consider a small deviation from the Einstein's theory.
Later in Sec.~\ref{SectionHS} we shall lift this approximation and vary the value of $R(0)$ to systematically find the one that better satisfies the boundary conditions.

The essence of the shooting method is to integrate the system of equations several times. With any starting guess of  $B(0)$, Eq.~(\ref{Matinfinity}) will normally not be satisfied, but the deviation of the obtained solution will serve to correct and provide an improved guess for $B(0)$. The procedure is thus iterated until convergence to a value of $B(0)$ that yields the boundary value of Eq.~(\ref{Matinfinity}) to a sufficiently good approximation. To control the error in the procedure we introduce 
(for intermediate computations only)
a $\beta$ parameter by
\begin{eqnarray}\label{29}
M(\beta,r)=\beta \, r+M(r),
\end{eqnarray} 
that vanishes when Eq.~(\ref{Matinfinity}) holds. The chosen linear dependence in $r$ of the spurious term is motivated empirically by numerous computer runs. In addition, $M(r)\propto r$ is the largest possible growth of $r$ compatible with a finite value of $\lim_{r\to\infty} B(r)$, as per Eq.~(\ref{24}) guaranteeing the asymptotic matching to a Minkowski solution.

\begin{figure*} 
 \begin{center}
 \includegraphics[width=0.3295\textwidth]{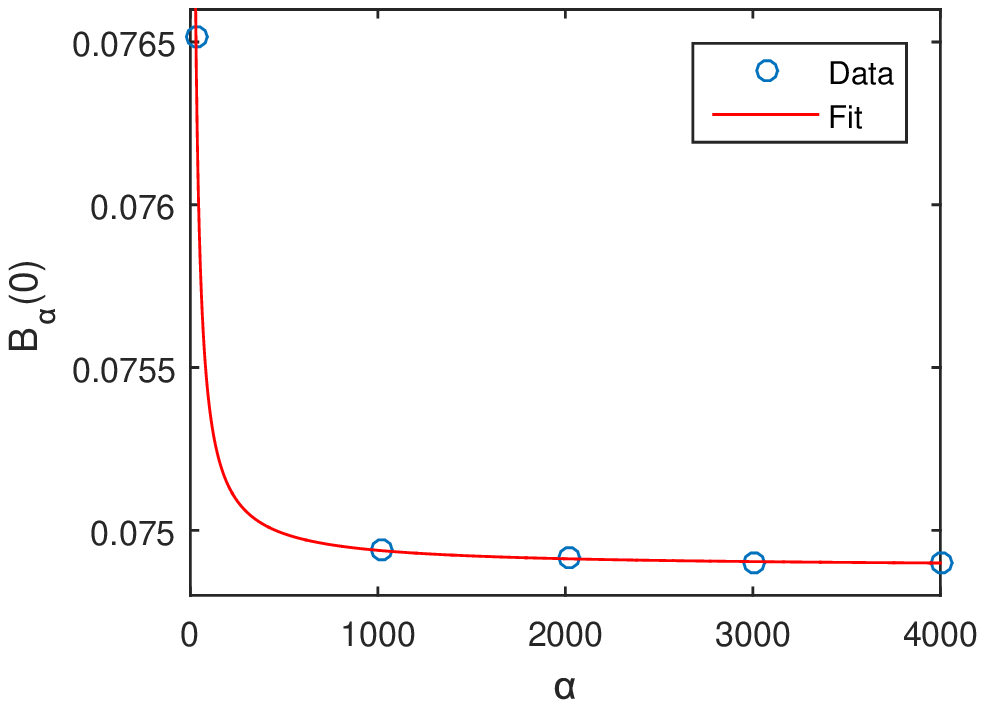} 
  \includegraphics[width=0.3295\textwidth]{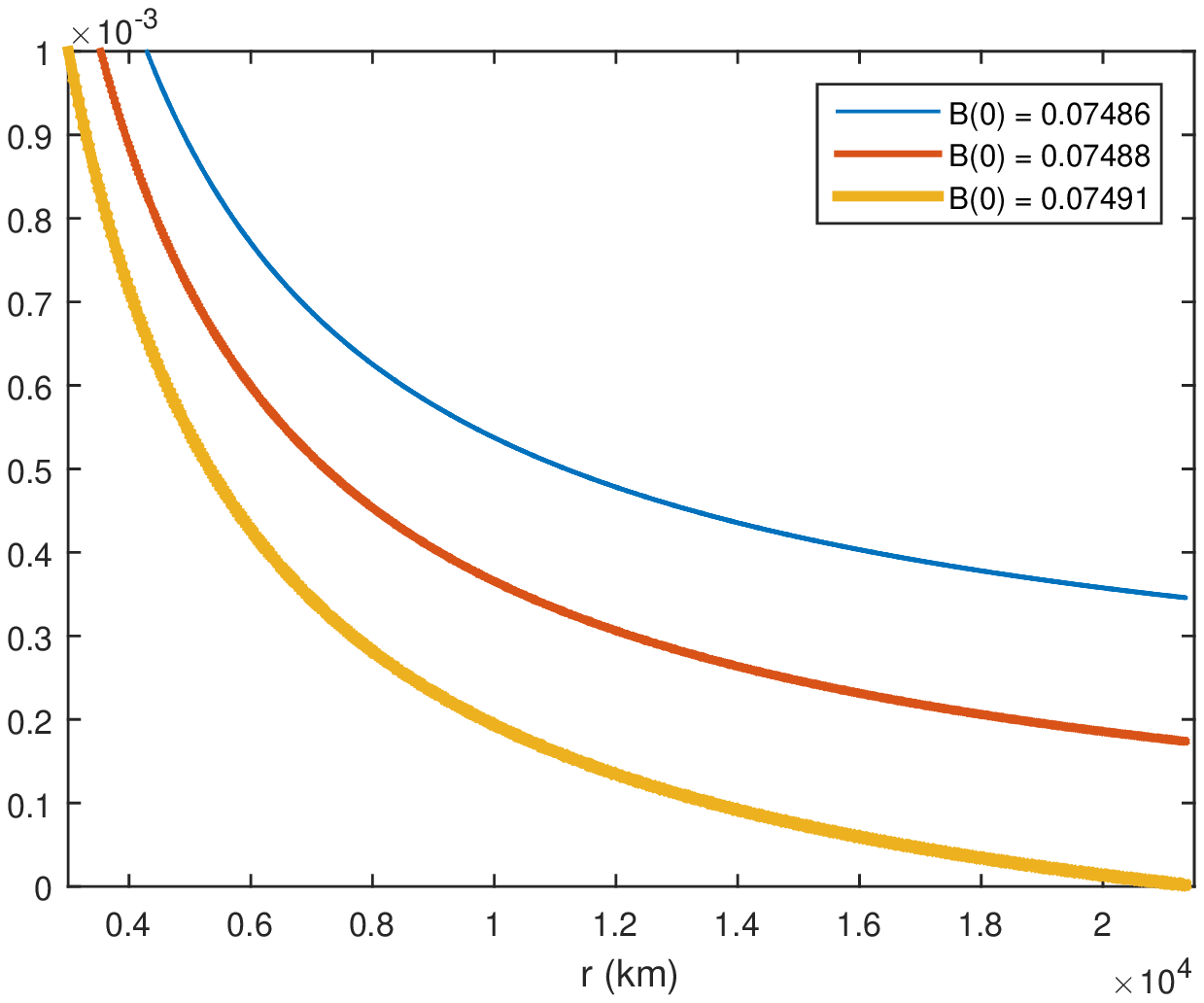}
 \includegraphics[width=0.3295\textwidth]{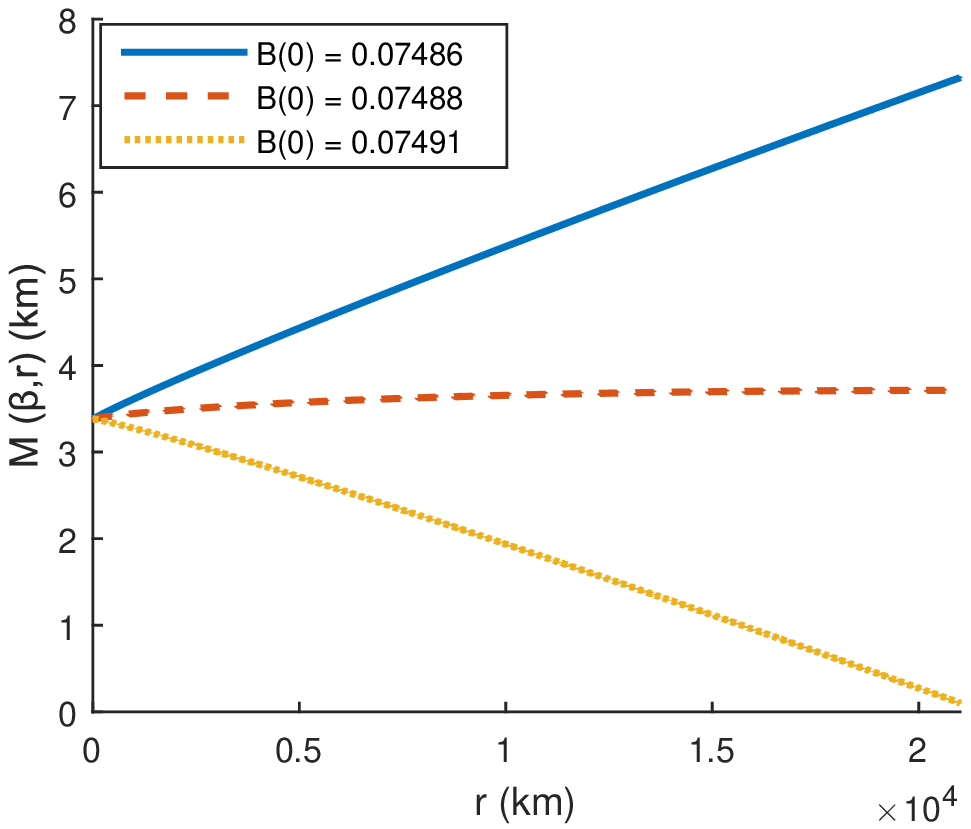}
  \end{center}
  \caption{
  Left panel: Initial value $B_{\alpha}(0)\equiv B_{\alpha r_\odot}(0)$ as a function of parameter $\alpha$ that shoots correctly to the boundary condition at $r=\alpha r_\odot$. Extrapolating $\alpha\to\infty$ we obtain $B_{\infty}(0)=0.7448(8)$.  
  Mid panel: we represent $M(r)/r$ for $B(0)$ extrapolated to satisfy Eq.~(\ref{Matinfinity}) in the central line, and for the two surrounding lines indicating the uncertainty that we have accepted in extrapolating $\alpha r_\odot\to \infty$ for this particular computer run.
  Right panel: $M(r)$.
}
 \label{Figure_1_wide} 
 \end{figure*}

Thus, we proceed to identify and minimise the contribution of $\beta$ in Eq. (\ref{29}) in order to find 
the $M(r)$ solution. 
The star's radius being $r_{\odot}$, we choose a sufficiently distant reference point in the star's exterior, $\alpha r_{\odot}$. 
For each $\alpha$ (again an auxiliary, intermediate parameter with no physical impact)
we choose this $B_{\alpha r_\odot}(0)$ by imposing that, at the reference point,  
\begin{eqnarray}\label{condbalpha0}
\left.\frac{M(\beta,r)}{r}\right\vert_{r=\alpha r_{\odot} } =0\ ,
\end{eqnarray}
which is the condition that we would like to satisfy at $r\to \infty$ as per Eq.~(\ref{Matinfinity}).
From the ansatz in Eq.~(\ref{29}), this determines $\beta$ as
\begin{eqnarray} \label{defalpha}
\beta+\frac{M(\alpha \, r_{\odot})}{\alpha \, r_{\odot}}&=&0\ .
\end{eqnarray}
Then we consider larger and larger values of $\alpha$ with the aim of reaching the limit $\alpha\to \infty$ as closely as possible in a finite computer.
The resulting $B_{\alpha r_\odot}(0)$ as function of $\alpha$ is plotted in Fig.~\ref{Figure_1_wide} left panel.
To extrapolate the computer data to $\alpha\to\infty$ (i.e., $r\to\infty$) we use a fitting function of the form $a_1 + a_2/\alpha^{a_3}$ with the $a_i$ being fitting constants\footnote{This generic form will be continuously used throughout the rest of the article to extrapolate any quantity to infinity.}.

		%
  %
%
When the dust settles, we have obtained  $B_{\alpha r_\odot \to \infty}(r=0)$ which is the initial value satisfying Eq.~(\ref{Matinfinity}). 
The numerical solution for the four involved functions $A(r)$, $B(r)$, $R(r)$ and $p(r<r_\odot)$ is now at hand and both initial and boundary conditions are satisfied. 

At this stage we turn to assigning a correct mass label for every physical configuration found.
In Fig.~\ref{Figure_1_wide}  we represent  the function $M(r)/r$ (where $M$ would be a constant in GR) for three values of $B(0)$ which represent the typical uncertainty in our numerical extrapolation of $\alpha\to\infty$ described above. In principle this uncertainty can be arbitrarily small provided longer computer time is allowed. 
When a calculation run has concluded, we assess how far in the radial coordinate 
the $M(r)$ function can be trusted  by comparing the terms in Eq.~(\ref{29}). The artifact $\beta r$ term, growing with $r$, will end up dominating for large enough $r$ so we have to extract the value of $M$ before then\footnote{This is akin to looking for a stability plateau in the extraction of an observable in lattice gauge theory before the noise overcomes the computation.}.
 
%
%
  %
%

\begin{figure} 
  	\includegraphics[width=0.480\textwidth]{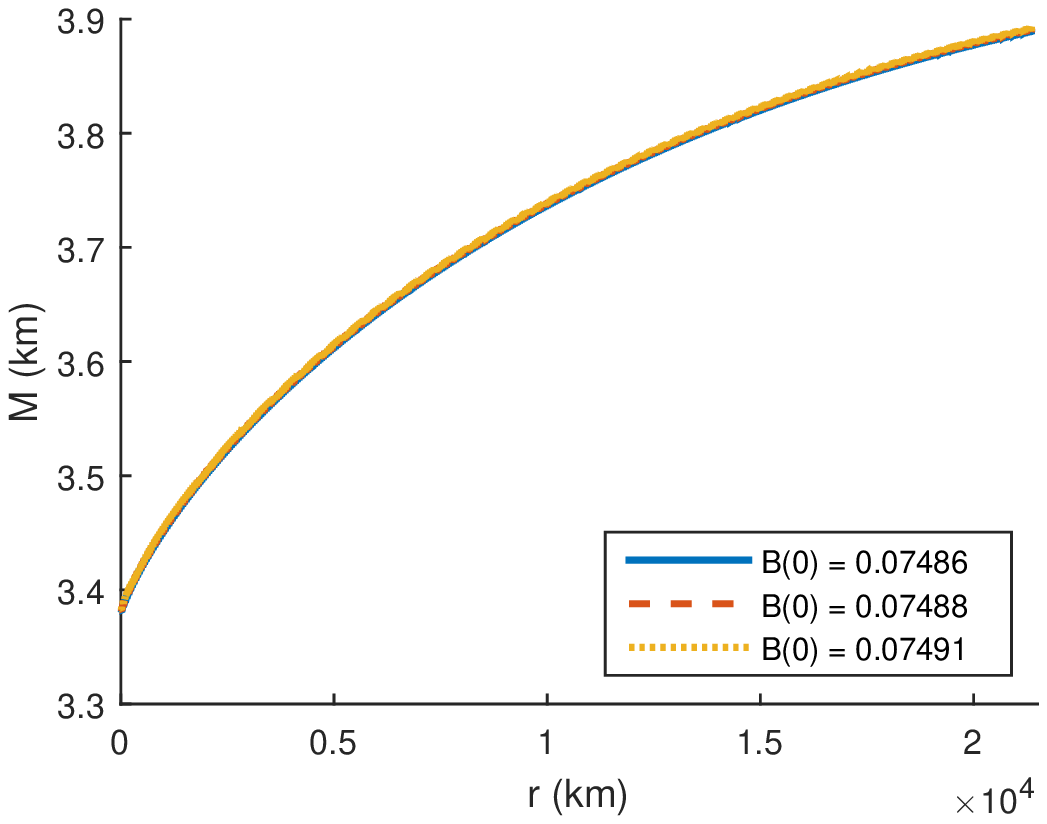}
	  	\includegraphics[width=0.480\textwidth]{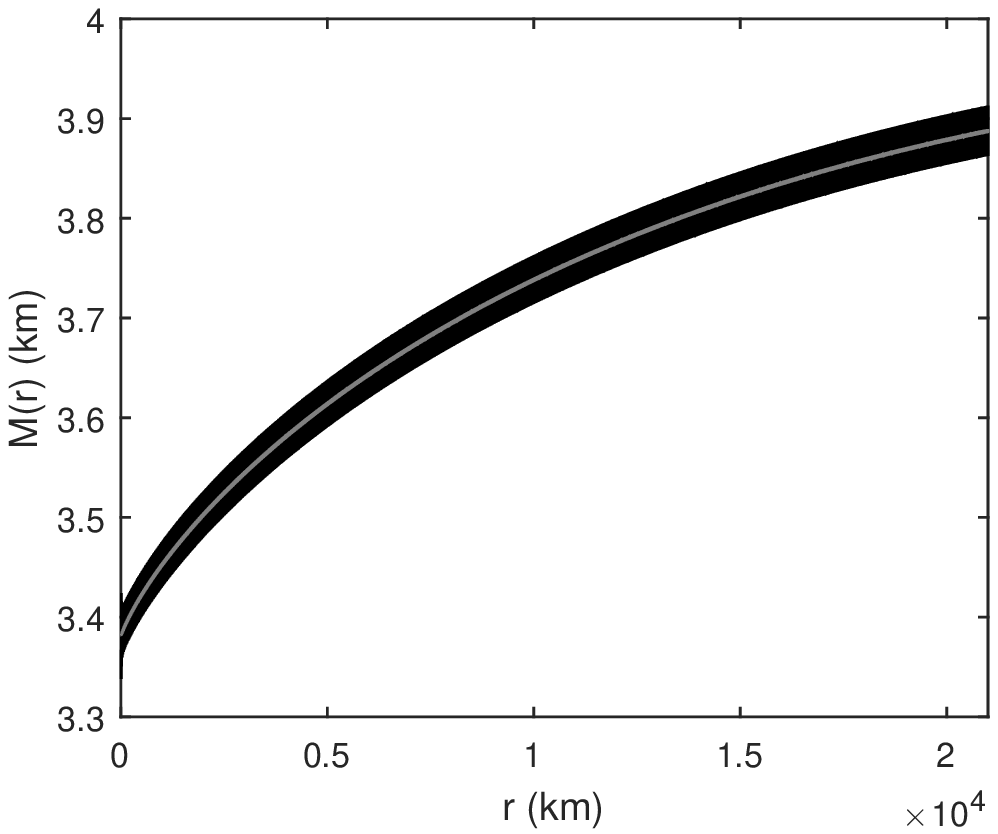}
		\caption{\footnotesize{ 
Upper panel: $M(r)$ once  the $\beta$ dependence has been extracted off $M(\beta,r)$. Since the three lines fall on top of each other, the error in the $\alpha r_\odot$ extrapolation is now negligible out to $r\sim 10^4r_\odot$. The relative error is less than $0.1\,\%$. (The $\alpha$ parameter of
Eq.~(\ref{defalpha}) has been taken here and in the following computations up to 2000). Lower panel: the black band represents the gravitational mass from Eq.~(\ref{27}), including the $U(r)$ factor (in this section we have not forced the constraint $U\to 1$ for large $r$ for the sake of expedience). Grey line: computed $M(r)$. As can be seen, the error is under control and can be easily quantified for each $a$ if need arises.
}}
  \label{Figure_3}
\end{figure}

In the example that we are showing, we kept enough precision so that the exterior solution can be trusted to radii $r=1000r_{\odot}$. For longer distances, when the radial coordinate becomes of order  $10^{4}$ km, the linear term catches up and dominates, as plotted in the right panel of Fig.~\ref{Figure_1_wide}. Thereafter the raw solution cannot be used anymore. 
However, we can still linearly fit $M(\beta,r)/r$ in all three cases depicted in Fig.~\ref{Figure_1_wide} central and right panels to obtain their respective values of $\beta$ and subtract the $\beta r$ term in Eq.~(\ref{29}), which increases  the numerical range of validity by another order of magnitude. The resulting  $M(r)$ satisfies the field equations by construction (all the intermediate solutions that we have found do so) and finally, the boundary condition~(\ref{28}) by extrapolation. With this extensive numerical effort 
we can satisfactorily find the exterior solution to $r\sim 10^4 r_\odot$ as finally displayed in Fig.~\ref{Figure_3} upper panel. 
 
As seen there, the three lines in Fig.~\ref{Figure_3} upper panel are almost indistinguishable, so that the extrapolation error in $M(r)$ can now be ignored. Let us point out that a small extrapolation uncertainty in $B(0)$ translates into a huge uncertainty (for large enough radii) for the intermediate function $M(\beta,r)$ but much less  for the correctly subtracted $M(r)$.
Moreover, the above panel in Fig.~\ref{Figure_3} clearly shows that the $M(r)$ function converges to a finite quantity as $r\to\infty$. This convergence was not a 
priori guaranteed due to the eventual appearance of residual, non-integer powers smaller than 1. These, by evading the boundary condition given by Eq.~(\ref{Matinfinity}), would have prevented the matching to a Schwarzschild solution at large distances.
We can then interpret the far solution as a Schwarzschild space-time and extract the observed mass $M_{f(R)}^{\infty}$.

\begin{figure} [H] 
  	\includegraphics[width=0.4975\textwidth]{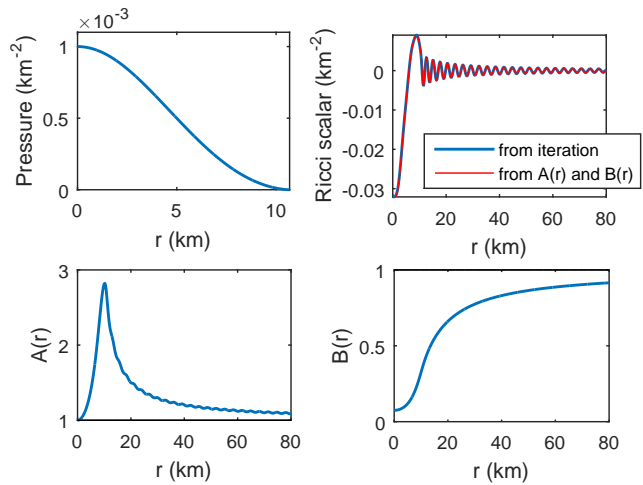}
		\caption{\footnotesize{Solution to the field equations. Clockwise from top left, $p(R)$, $R(r)$, $B(r)$ and $A(r)$.
The curvature scalar $R(r)$ performs damped oscillations, also seen in $A(r)$ (and present in $U(r)$, not shown here). The oscillation of $B(r)$ is much smaller, and can barely be appreciated in this example. The iterated and direct (from $A(r)$ and $B(r)$) computation of the curvature scalar $R(r)$ agree very well within our numerical precision.
The top right panel shows a very clear standing wave of the Ricci scalar outside the $f(R)$ star.
}}
  \label{Figure_5a}
\end{figure}

Continuing with the analysis of systematic uncertainties, let us briefly  return to address the initial approximation $R(0)=\kappa\,T(0)$.  
With this {\it a priori} assumption, we were not satisfying Eq.~(\ref{28}), only approximately, so that
 $U(\infty)\! \not=0$ and thus $A(\infty)\!\not=1$. 
It is then convenient to compare the $M(r)$ function with the actual $M_{f(R)}$ value from Eq.~(\ref{27}), the expected ratio being the $(1+U(r))$ factor.
This comparison is carried out in Fig.~\ref{Figure_3} lower panel.
%
		%
  %
%
The only noticeable effect is that, since $U(r)$ is oscillating (as can be deduced from $A(r)$ in Fig.~\ref{Figure_5a}, the mass function $M_{f(R)}$ has small oscillations that are rather irrelevant at large distances. 
Therefore, we have checked that the error induced by simplifying the initial condition for $R(0)$ 
remains under control. This adds up to the error in the finiteness of the maximum $\alpha r_\odot$ where the boundary condition on $M(r)/r$ is applied, but both are small for the values of $a$ under study. 

In order to wrap up the discussion of this section, we represent in Fig.~\ref{Figure_5a} the obtained metric coefficients and the curvature scalar in this example. These plots provide a useful check of the computer codes. 
Until now we have solved for  $R(r)$  in the $f(R)$ theory following the dynamical Eq.~(\ref{8}). But in the end it can also be computed from the metric functions $A(r)$, $B(r)$ and their derivatives via the Christoffel symbols and the Ricci tensor\footnote{This was not possible ahead because $A$ and $B$ had not been previously solved for.}, as encoded in Eq.~(\ref{10a}), and we can compare both to check self-consistency.  The perfect agreement between the two methods  is plotted in the upper right panel of Fig.~\ref{Figure_5a}.

\section{$f(R)=a\,R^{2}$ Model }\label{SectionRsq}
The quadratic model $f(R)=a R^2$ has attracted intense attention in the last years in the context of astrophysical applications, as well as having been widely used to provide a geometric origin to dark matter abundances \cite{Cembranos:2008gj}. It has also been recently rediscovered to explain cosmic inflation \cite{staro, Planck-Inflation}. 
One would naively expect solar system tests of GR to impose tight constraints on the parameter $a$, but this is not quite so. What the tests indicate is that, broadly speaking, 
corrections to the Einstein-Hilbert gravitational Lagrangian on those scales are very small, $\arrowvert f(R_{ss})\arrowvert <10^{-6}$  ({\it c.f.} constraints on several cosmological scales in \cite{Lombriser:2011zw}).
Nevertheless one needs to bear in mind that the possible curvatures are themselves very small even inside the Sun, so that even if the contribution $aR^2$ is small, $a$ might still be sizeable. 
Actually in this context the performed tests are exterior-metric tests where $R_{ss}=0$ exactly in GR. Of course, in extended theories of gravity, $R$ might not be zero. In fact, its natural size can be estimated
from the Kretschmann scalar $R^{\mu\nu\rho\sigma}R_{\mu\nu\rho\sigma}= 12 r_S^2/r^6$ (with $r_S$ the Schwarzschild radius) that does not vanish in GR and, at $r=r_\odot$ has a value of about $3\times 10^{-17}{\rm km}^{-2}$.
This means that $a$ is very much unconstrained by such tests\footnote{
%
%
The value for the Starobinsky inflationary model in good agreement with Planck latest results can be constrained using the 
CMB anisotropies  $a\sim 10^9 M_P^{-2}$ ({\it c.f.} Sec. 2 in \cite{Riotto}).  Thus  
the inflationary value of $a$ turns out to be $a\sim 2.74\cdot 10^{-49} {\rm km}^{2}$. Consequently, for the Ricci curvatures of interest in our problem, this value would provide negligible corrections to the standard Einstein-Hilbert term. 
Thus, the quadratic models presented in the following must be understood as GR effective corrections which appear at gravitational curvatures such as those characterising the neutron stars dynamics.
} and any sizeable value would have an important effect in much more unexplored neutron star physics given the large field intensities.

\begin{figure*} 
 \begin{center}
 \includegraphics[width=0.3275\textwidth]{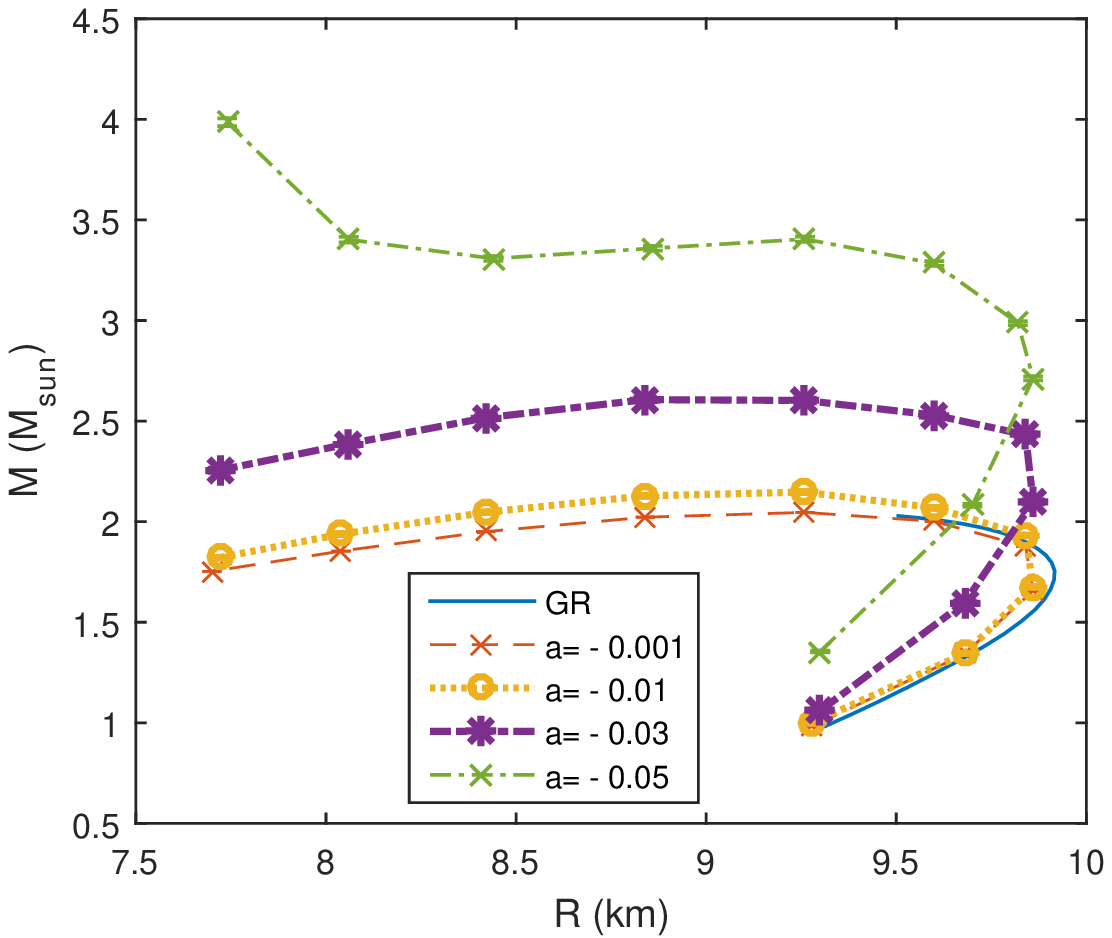} 
  \includegraphics[width=0.3275\textwidth]{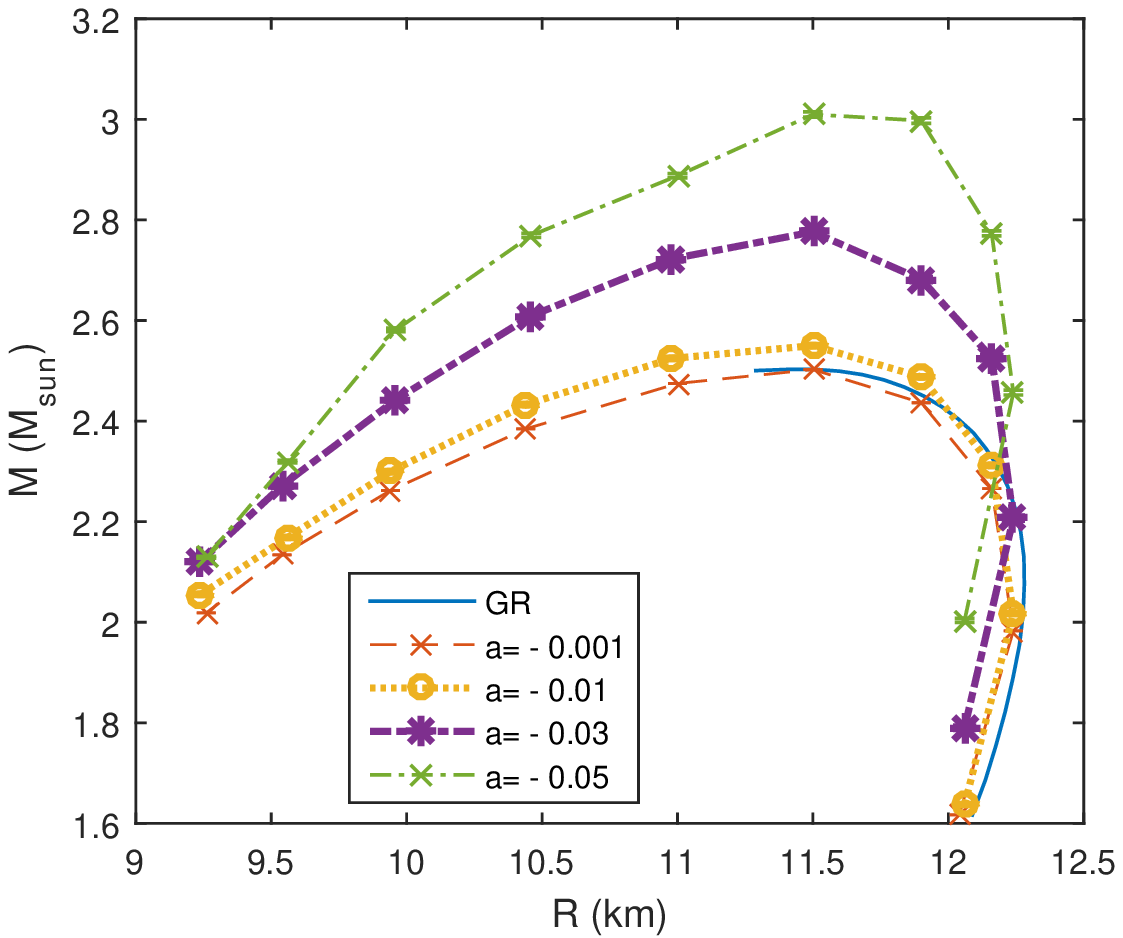}
 \includegraphics[width=0.3275\textwidth]{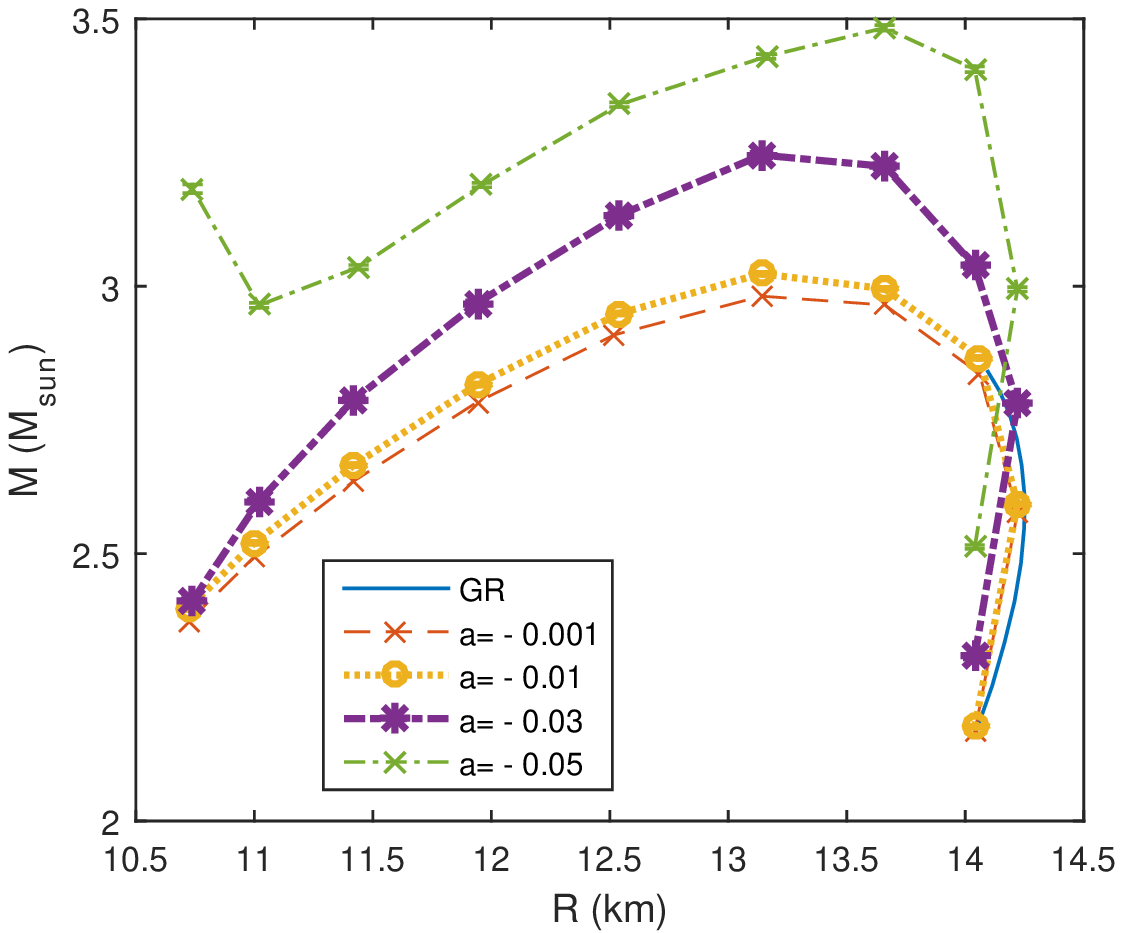}
  \end{center}
  \caption{Mass-Radius diagram for different equations of state in the $f(R)=aR^2$ model: Soft (left), Middle (centre) and Stiff (right). In all the panels, the GR Mass-Radius diagram has been included for comparison. As seen, for the same values of the parameter $a$, 
masses are larger and radii are smaller than in GR. Both with the Softer and Stiffer EOS, solutions with high masses are found for small radii, here shown for the case $a=-0.05$. For bigger values of $|a|$, the diagrams for all the EoS would provide higher and higher masses. The points correspond to $\{R,M\}$ pairs obtained through numerical simulations. (Points: outcome of calculation. Lines: simple interpolation.)}
 \label{Figure_Mass_Radius_Diagram} 
 \end{figure*}

As derived from Eq.~(\ref{20}), the parameter $a$ must be negative in order to guarantee damped oscillatory solutions. By making use of the numerical analysis detailed above, in this section we shall obtain star configurations in a wide range of central pressures for the EoS presented in Fig. \ref{Figure_5}. 
For each EoS the mass-radius curves will be obtained for different values of the parameter $a$ as depicted in Fig. \ref{Figure_Mass_Radius_Diagram}.
As seen in all of them, the mass is an increasing function of $|a|$ obtaining for each radius a mass larger than the GR counterpart, this effect being more appreciable for 
larger values of $|a|$. This phenomenological trend can be understood from inspection of the numerical solutions. Indeed, the parameter $M(r)$ turns out to be increasing with the radial coordinate till  at sufficient distance away from the star, it reaches a limiting value $M_{f(R)}^{\infty}$. 
This behavior is explained by the damped oscillations that the Ricci scalar experiences outside of the star. As previously shown in Fig. \ref{Figure_5a}, the Ricci scalar is not zero in the outer region, but asymptotically tends to this value at sufficient distance from the star. This oscillatory effect is more important for bigger values of $|a|$ and can thus be interpreted as the presence of an effective content of energy (or matter) outside the star, this energy (or matter) component being the dominant source of the external gravitational field budget at  distant regions from the star.

Careful observation of Fig.~\ref{Figure_Mass_Radius_Diagram} will show the reader that 
the maximum achievable mass does not behave monotonically with the sound speed (the stiffness of the EoS): the ``middle'' equation of state actually produces a maximum stellar mass that is below \emph{both} the ``soft'' and the ``stiff'' variants by a significant amount, about half a solar mass. This phenomenon will reappear in the next section in Fig.~\ref{Figure_Mass_Radius_Diagram2}.

We have pursued this feature, with an explanation provided in Fig.~\ref{fig:resonance} in the next section; it can be attributed to the oscillatory nature of the system of dynamical equations, that can give rise to resonance-like phenomena at the star radius, and to the observed large contribution to the apparent star mass from outside the star.

In conclusion of this section, stable $R+aR^2$ models yield star masses that are generically much larger than in Einstein's theory and finding further two-solar mass stars, or even larger ones, does not  constrain this class of models. On the other hand, for a given mass, the radii are significantly smaller than in GR (because much of the mass seen from the observer at infinity is actually external field energy density) which can turn out to be phenomenologically relevant.

\section{Hu-Sawicki model}
\label{SectionHS}
In this section we analyze  a class of $f(R)$ models, the so-called Hu-Sawicki model~\cite{Hu:2007nk}, whose functional form is 
\be
f(R)=-cH_0^2\frac{b(R/cH_0^2)^n}{e(R/cH_0^2)^n+1}\,,
\label{1.3}
\ee
which for a suitable choice of the parameter space is able to provide late-time cosmological acceleration, so it has drawn a vast attention from the specialised community  \cite{DombrizPRD2016}.
Its parameters are the dimensionless $\{b,c,e\}$ constants, and $H_0$ is the Hubble expansion parameter. The integer $n$ chooses among several different model variations, and for the sake of simplicity we shall focus on the $n=1$ case. Thus a simple reparameterisation takes Eq.~(\ref{1.3}) to
\begin{eqnarray}\label{30}
f(R)=-\frac{b\,R}{1+b\,\frac{R}{d}},
\end{eqnarray}
where $b$ is again an adimensional parameter;  $d$ has now dimensions $[L^{-2}]$,
as does the curvature scalar, and there is a model watershed at $R\simeq d$.

Additionally, Eq.~(\ref{20}) produces the stability parameter
\begin{eqnarray}\label{31}
\gamma=\frac{A}{6b^{2}}\,d\,\left(1-b\right),
\end{eqnarray}
that must be negative to yield solutions asymptotically compatible with a flat spacetime at $r\to \infty$. This depends on the signs of $d$ and $(1-b)$.
Thus we have analysed separately the cases  $|R|>|d|$ and $|R|<|d|$, with both $b<1$ and $b>1$, in the following. 

\subsection{$|R|>|d|$}\label{Rmayor}

If we expand Eq.~(\ref{30}) with $d/R$ as the small parameter we obtain
\begin{eqnarray}\label{32}
f(R)\simeq -d + \frac{d^2}{b R}\ .
\end{eqnarray}
The dominant term in this expression is a cosmological constant ($-d$),
which is immediately visible in the solution spacetimes.
To illustrate it, we have considered $d=10^{-3} {\rm km}^{-2}$ and $b=2.5$, whose substitution in Eq.~(\ref{31}) satisfies the stability condition  $\gamma<0$.

\begin{figure} [H] 
  	\includegraphics[width=0.49750\textwidth]{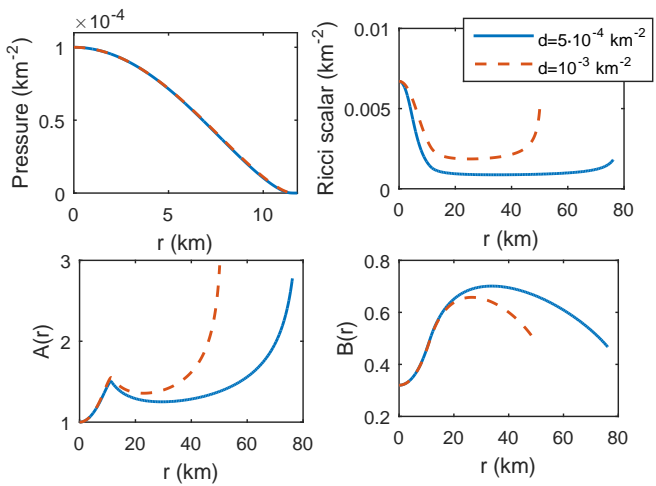}
		\caption{\footnotesize{
Hu-Sawicki solutions with $b=2.5$, both with the mid-stiffness EOS and the same initial condition,
but two values of $d$ (determining the cosmological constant). 
The exterior solution starts at $11.72\, {\rm km}$ where pressure vanishes. 
Both solutions are of Schwarzschild-de Sitter type;  a smaller cosmological constant entails a further cosmological horizon.
}}
  \label{Figure_9}
\end{figure}

With such appreciable values of $d$ we recover a Schwarzschild-de Sitter spacetime as shown in figure~\ref{Figure_9}. This spacetime produces a cosmological horizon at $\mathcal{O}(10\ r_\odot)$, and 
hinders us from taking the $r\to\infty$ limit.
 This is not an admissible spacetime for stellar applications in astrophysics, since the cosmological constant is in blunt disagreement with observation, and we need to ignore such parameter values.

Instead, if we take a value of $d\simeq  \Lambda$ that matches the present-day cosmological constant, the contribution of the $f(R)$ in Eq.~(\ref{32}) to the total action is totally negligible, so that we recover the standard GR Schwarzschild spacetime case. Anyhow, 
we expect this feature to persist in other similar $f(R)$ theories that remain viable in cosmology \cite{Other_f(R)}.
%
%
%
Therefore this case with $|R|>|d|$ is either at odds with observation (large cosmological constant) or trivial (too closely Schwarzschild). In view of it, we proceed to examine the second possibility.

\subsection{$|R|<|d|$}\label{Rmenor}

We then return to Eq.~(\ref{30}), considering now an expansion in $\frac{R}{d}$ and obtaining
\begin{eqnarray}\label{33}
f(R)\simeq-b\frac{R}{d}+\frac{b^{2}}{d}\left(\frac{R}{d}\right)^{2}\!\!  -\frac{b^{3}}{d^{2}}\left(\frac{R}{d}\right)^{3}+\mathcal{O}\left(\frac{R}{d}\right)^{4},
\end{eqnarray}
naturally assuming that $f(R)$ is analytic around $R=0$.
There are two parameter swaths compatible with the stability criterion in Eq.~(\ref{31}),
which is a product of two signed numbers; let us visit them in turn.
\paragraph{$d>R>0,   \,\,\,   b>1$}

\begin{figure} 
  	\includegraphics[width=0.4975\textwidth]{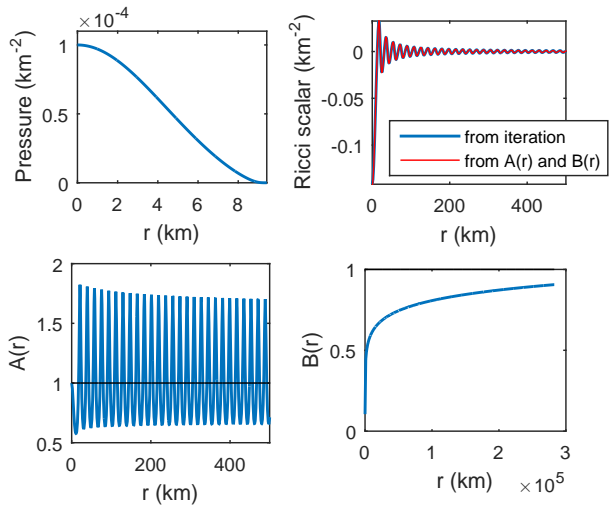}
		\caption{\footnotesize{
Static, spherically symmetric spacetime in Hu-Sawicki model with $b=12.5$ and $d=10\,{\rm km}^{-2}$. 
This case satisfies  $R<d$ as the curvature scalar is of order $0.1\,{\rm km}^{-2}$. 
Because General Relativity is substantially modified, the gravitational mass is a huge
 $10^{4}\,M_{\rm Sun}$, and this parameter combination is presently of no astrophysical use.
Nevertheless it shows that in theories beyond GR, very large neutron star masses are possible.}}
  \label{Figure_10}
\end{figure}

This parameter combination cannot be continuously connected with the GR case
in which Eq.~\ref{30} yields $f(R)=R$ since we need to maintain $R<d$ yet $b$ is positive.
This entails that the resulting spacetimes cannot be guaranteed to parametrically and perturbatively deform to the Schwarzschild solution of GR (and in fact they do not, as shown in Fig.~\ref{Figure_10}). Even if the solutions are very different from GR's, we may still analyze them numerically with our developed methods. The star masses turn out much larger than in GR since, as observed in  Fig.~\ref{Figure_10}, the $B(r)$ function grows much more slowly than in the GR Schwarzschild case.
Moreover, the function $A(r)$ oscillates with an amplitude larger than, for example, the $aR^2$ case displayed in   Fig. \ref{Figure_5a}, where modifications from GR were kept small. 
In brief, the solutions are phenomenologically similar to those in the  $a\,R^{2}$ model, except that, since the modifications from General Relativity are much larger, the effects of $f(R)$ are amplified.

\paragraph{$d<0,   \,\,\,   b<1$}

\begin{figure} [H] 
  	\includegraphics[width=0.4975\textwidth]{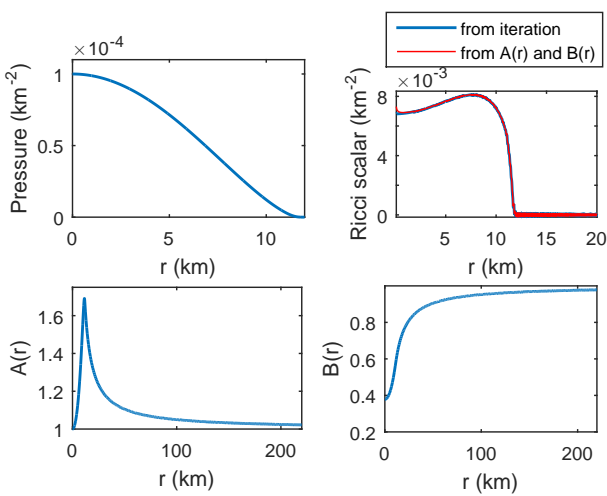}
		\caption{\footnotesize{Static, spherically symmetric Hu-Sawicki metric coefficients and scalar curvature with parameters $b=10^{-2}$ and $d=-1\,{\rm km}^{-2}$. The exterior Schwarzschild solution is asymptotically recovered.}}
  \label{Figure_11}
\end{figure}

This case does allow Eq.~(\ref{30}) to be matched to GR, which is achieved by letting $b\to 0$; we show an example of this case in Fig.~\ref{Figure_11}.
Taking as a starting point the solution of Einstein's theory, we increase stepwise the value of  $b$, obtaining the pertinent modification from  $f(R)$ separating us from GR.  The solutions obtained are of the same type as those in Fig.~\ref{Figure_5a} and, carrying out an analysis analogous to that in Sec.~\ref{sec:numerics}, 
we obtain masses for the various EoS, in the same way as in Sec.~\ref{SectionRsq}.
\begin{figure*} 
 \begin{center}
 \includegraphics[width=0.32975\textwidth]{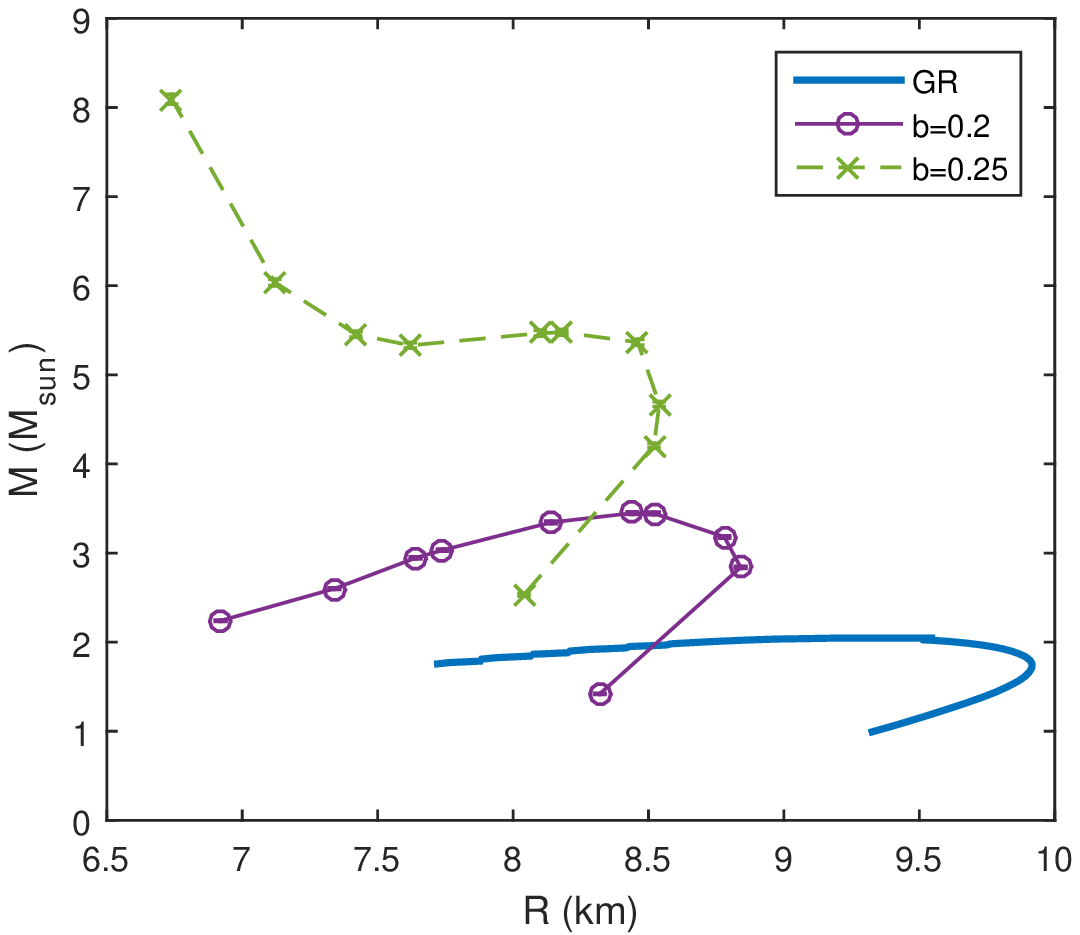} 
  \includegraphics[width=0.32975\textwidth]{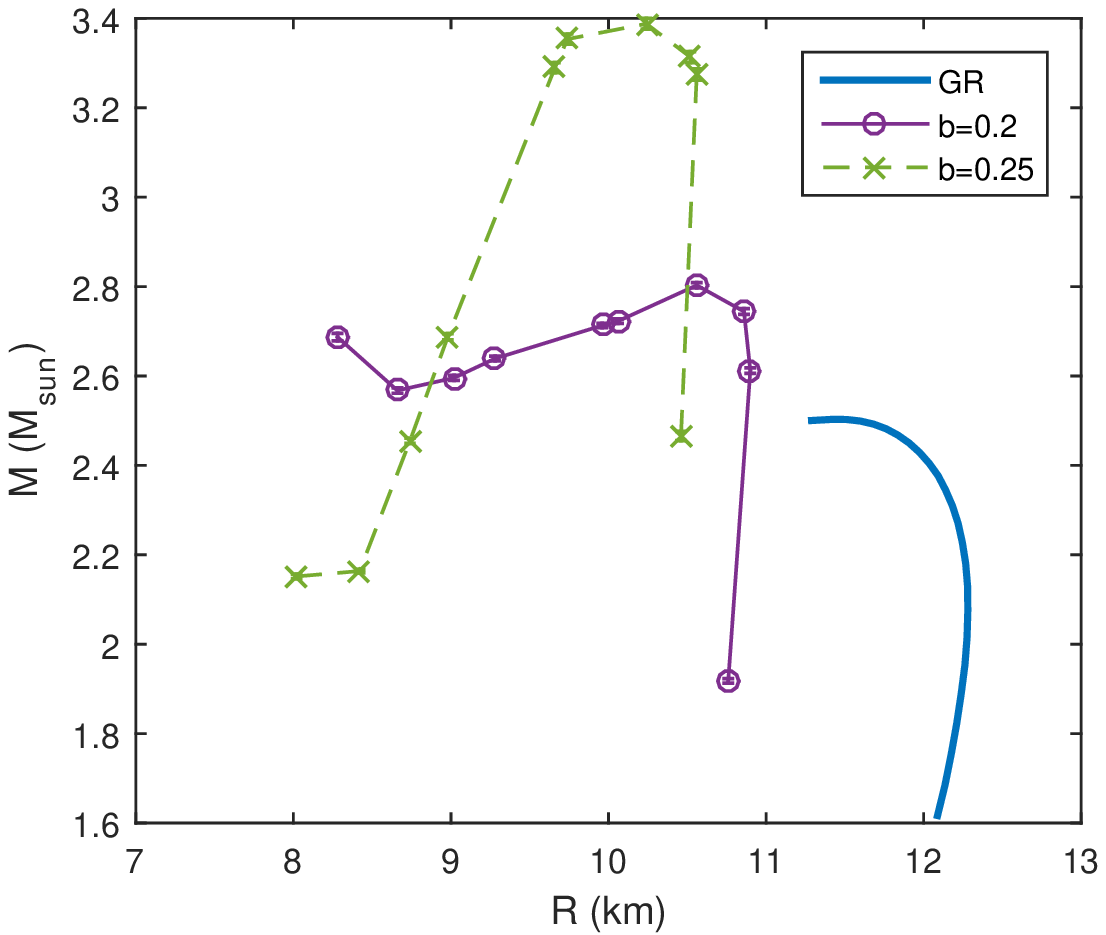}
 \includegraphics[width=0.32975\textwidth]{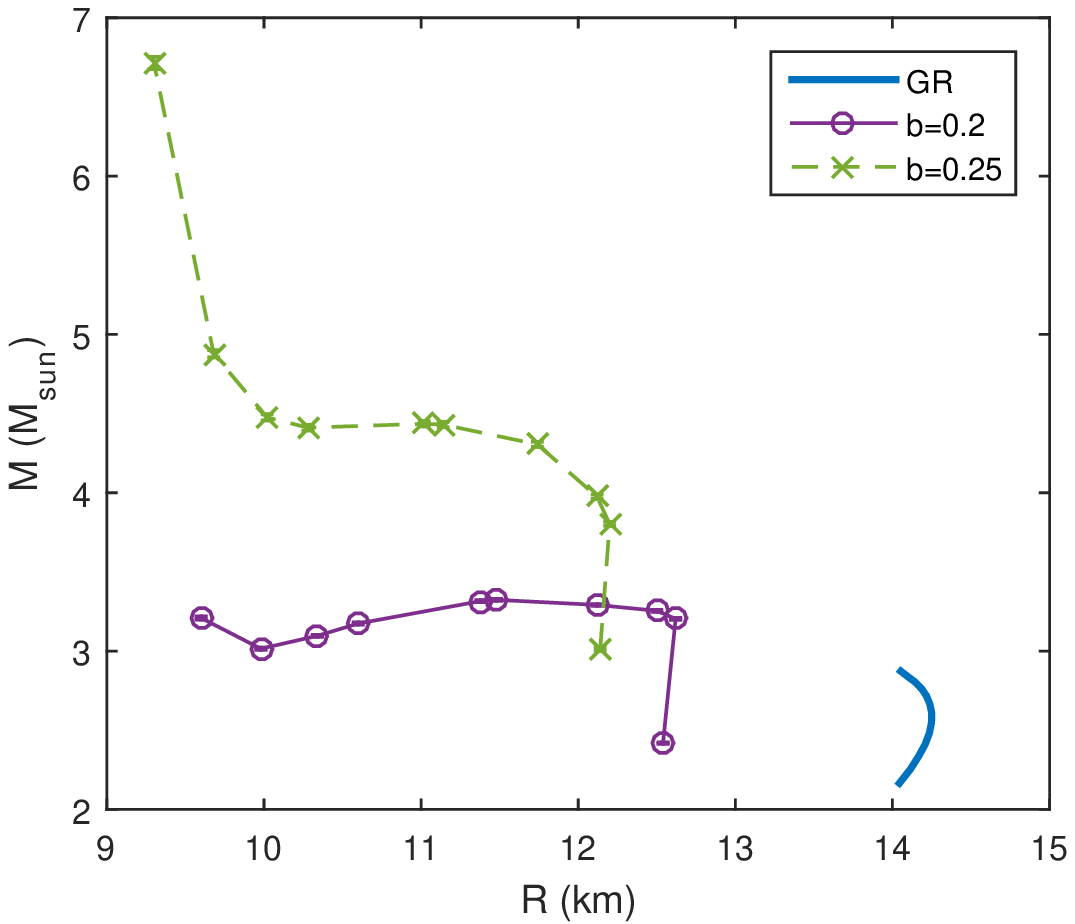}
  \end{center}
  \caption{Mass-radius diagram for neutron stars in Hu-Sawicki's $f(R)$ model. 
We fix $d=-1\,{\rm km}^{-2}$ and vary  $b$. 
From left to right we have employed the softer, middle, and stiffer equations of state from~\cite{Hebeler:2013nza}. General Relativity is shown in all plots as a benchmark.
With the small exception of the ``middle'' equation of state at very high pressures, we find that generally the masses increase with $b$.
Also important is to note that the radii are smaller than in GR (much of the mass is outside the star).
}
 \label{Figure_Mass_Radius_Diagram2} 
 \end{figure*}

\begin{figure}
\includegraphics*[width=1\columnwidth]{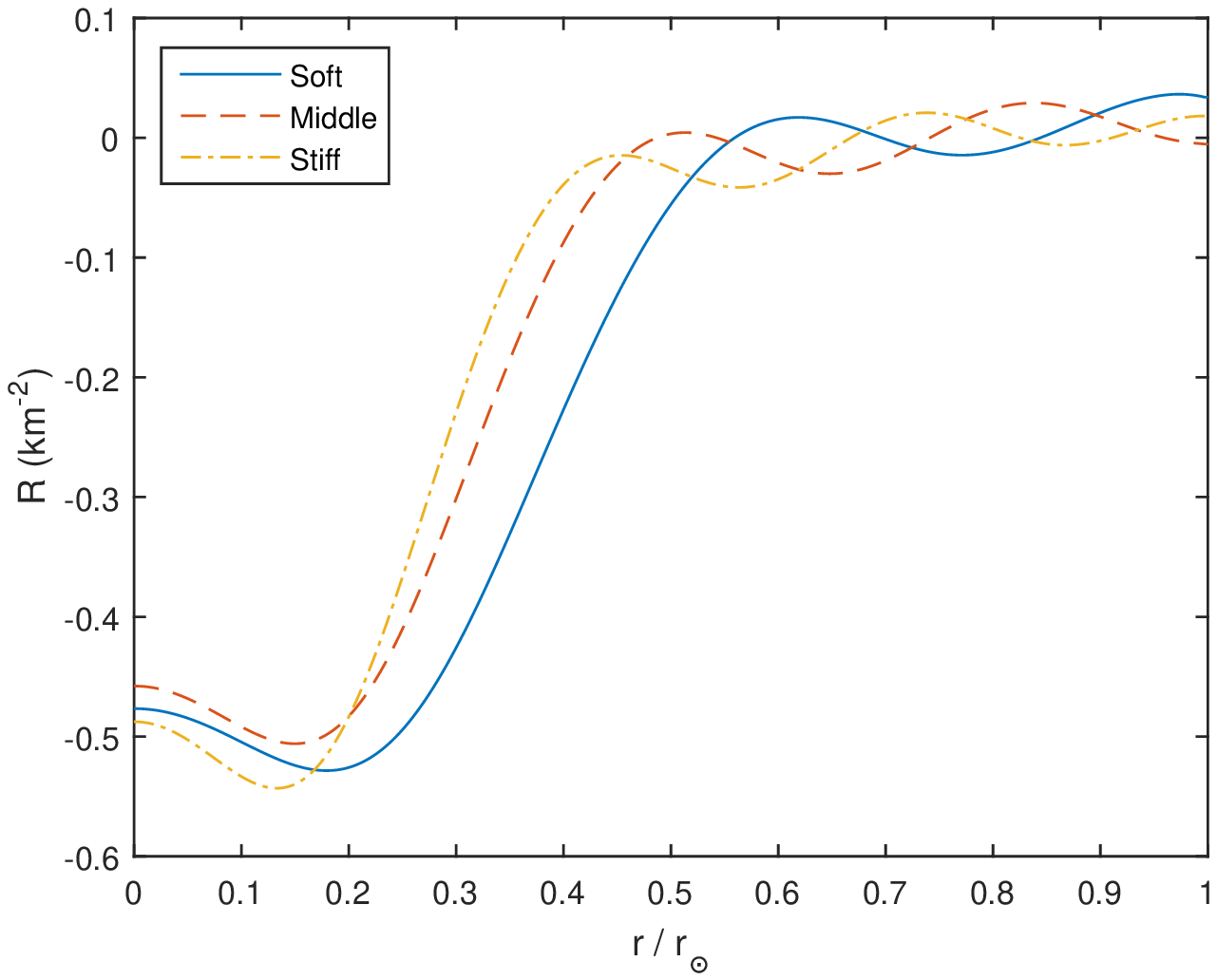} \\
\includegraphics*[width=1\columnwidth]{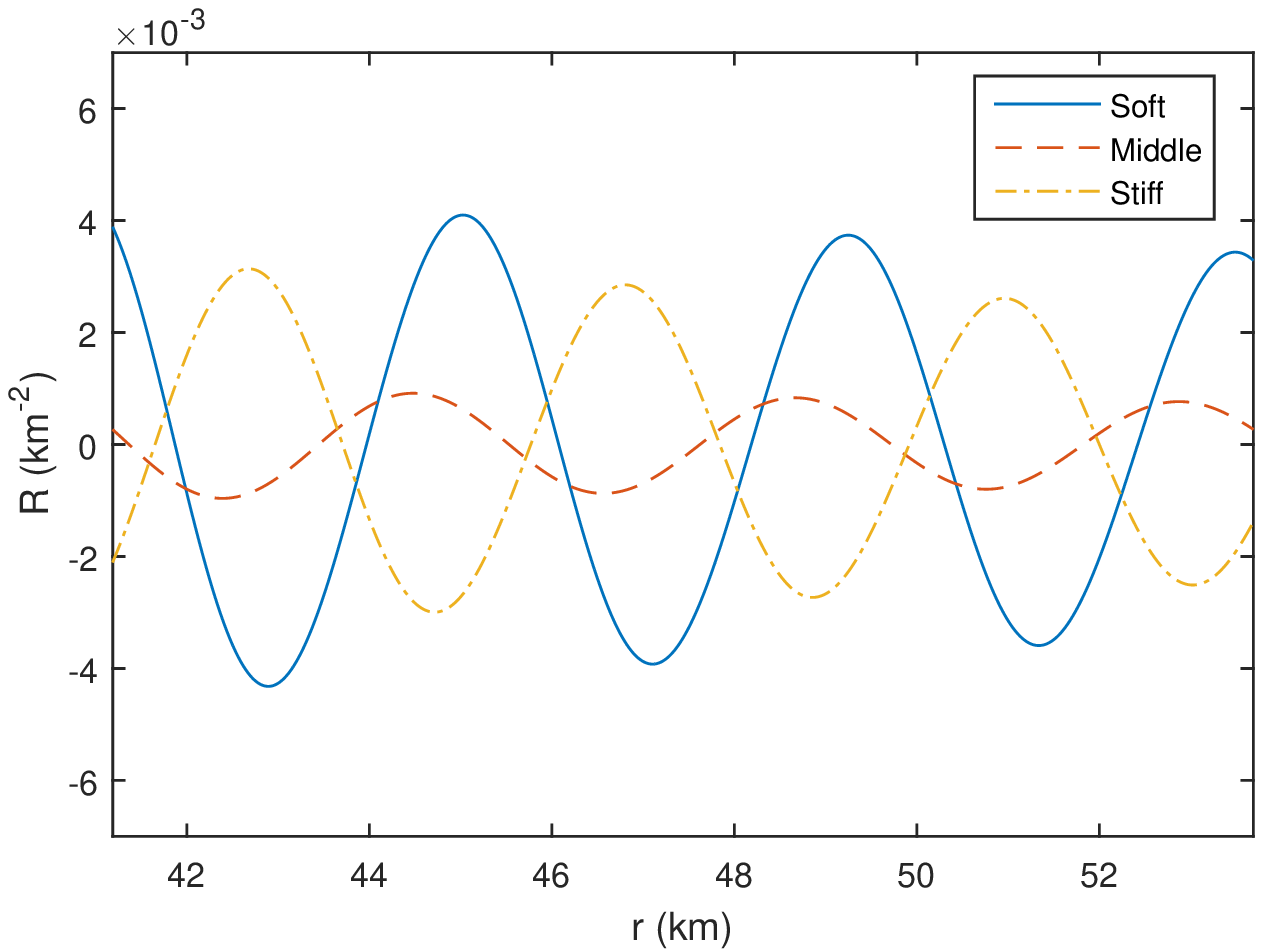}
\caption{\label{fig:resonance} 
Ricci scalar inside (upper panel) and outside but near (lower panel) a neutron star in Hu-Sawicki model
(same as in Fig.~\ref{Figure_Mass_Radius_Diagram2} with $b=0.25$ fixed). The central pressure is
$p=10^{-2}{\rm km}^{-2}$. The bottom plot shows that the ``middle'' EoS of Hebeler {\it et al.}~\cite{Hebeler:2013nza} has the smallest amplitude of (secularly damped) Ricci scalar oscillations. This explains that 
 the exterior metric when evaluated at $r\to\infty$ appears to have the smallest mass.
The top plot shows how this comes about: when the corresponding inner metric hits zero pressure, the star with the ``middle'' EoS happens to be at a minimum of the oscillation. Thus, the non-monotonic effect of $M$ with the EoS is due to the oscillating nature of Eq.~(\ref{5}-\ref{8}).}
\end{figure}

Once again,  Fig.~\ref{Figure_Mass_Radius_Diagram2} shows that
the maximum achievable mass  does not behave monotonically with the sound speed (or the EoS stiffness). In this case, the effect is even more remarkable than in the example of 
Fig.~\ref{Figure_Mass_Radius_Diagram}. We clearly see in Fig.~\ref{fig:resonance} how it is due to the oscillatory nature of the system of dynamical equations, and to the large contribution 
to the apparent star energy content from outside the star.

What happens is that this significant contribution to the Newtonian-matched star mass from the star's outer metric is suppressed for the ``middle'' EoS that presents a smaller oscillation of the Ricci scalar (bottom plot of fig.~\ref{fig:resonance}). 
This is because when the radius of the star ($P=0$) is reached from inside upon integrating the system of equations of Eq.~(\ref{5}-\ref{8}), which are oscillating, the ``stiff'' and ``soft'' EoS both happen to give an $R$ near a maximum of the oscillation, while the ``middle'' EoS happens to yield $R$ near a minimum. This is a typical wavelike phenomenon 
and it is thus not surprising after careful consideration.

Moreover, the only (fine-tuned) scenario where the outer metric can be Schwarzschild (with $R=0$) happens whenever $R\arrowvert_{r=r_\odot}=0$,
which corresponds to a resonance-like phenomenon fitting a standing $R$-wave precisely inside the star.
As we can see in Fig.~\ref{Figure_Mass_Radius_Diagram2}, 
the general trend is that the presence of $f(R)$ contribution in this case increases the maximum star
masses and reduces the radii.

\section{$f(R)$ gravitational waves effects in stars merger}
\label{sec:GW}

 The existence of gravitational waves in $f(R)$ theories has already been demonstrated~\cite{GW_f(R)}
 and a peculiarity thereof is the propagation of an additional longitudinal mode, surplus to the transverse ones in General Relativity. This corresponds to the propagation of the $\varphi$ scalar field in the canonically equivalent scalar-tensor modified gravity theories.

In any case, we have shown in this work that the \emph{quantity of matter} does not coincide with the \emph{total mass} of the neutron star given the large amount of energy-equivalent stored in the distorted external spacetime in the form of standing waves (see the scalar $R$ curvature in Fig.~\ref{Figure_5a}).

During a merger of two such compact objects, a large amount of this energy must be released. Since the outer space time solution is pure-gravity, this can be achieved in the form of gravitational waves. 
Nevertheless, the existence of an additional scalar mode in these theories would produce that the merger of the compact objects
also involves the emission of scalar waves. Since the latter would be sourced by the monopole of the system, whereas the gravitational waves are sourced by the quadrupole contribution, the straightforward prediction would be that the amount of energy emitted in the scalar mode might be considerably larger than the emission in GWs. Eventual conflicts of this effect with well-established observational results, such as binary pulsar observations, could be avoided thanks to the chameleon mechanism present in $f(R)$ theories \cite{Weltman}.

Anyhow, the time it takes to shake all the gravitational energy involved in the merging is surely large, and requires dedicated calculation to see if it can match the LIGO observations, but from our estimates above, achieving a deficit of several solar masses should be feasible. This is in contrast with General Relativity mergers of two neutron stars, that are at most about 2 solar masses each, and can only emit a fraction thereof in the form of gravitational waves. 
 At this stage, we must stress in agreement with the discussion in the previous paragraph that part of the released energy would be in form of the scalar mode.

\begin{figure}
\includegraphics[width=0.6\columnwidth]{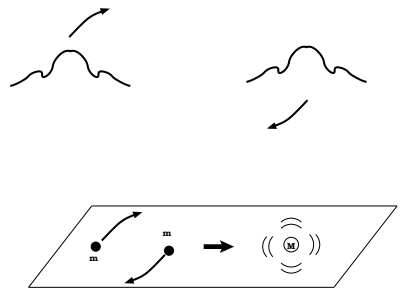}
\caption{\label{fig:radiacion}
Two compact stars in $f(R)$ theory, with a non-trivial exterior $R$ configuration
may be seen, at a large distance, as two merging masses of irrelevant inner structure;
because the $m$ that we have carefully defined and calculated in this article comes from
guaranteeing a Minkowski space at infinity, its emergent asymptotic Lorentz symmetry ensures
energy conservation, so that if the final state is a compact star of mass $M$, the radiated
energy can be of order $E=M-2m$. Thus, assuming a small leak into the scalar mode, there is room for ample gravitational wave radiation of order several solar masses given the mass-radius curves in figs.~\ref{Figure_Mass_Radius_Diagram}
and~\ref{Figure_Mass_Radius_Diagram2} that show how static neutron stars can store a lot
of gravitational energy in the outer field in $f(R)$.}
\end{figure}

Nonetheless, our estimates are relevant because our careful definition of the mass ensures that we have an asymptotic Minkowski space time very far from the neutron star, as illustrated in Fig.~\ref{fig:radiacion}. 
The Lorentz symmetry of the asymptotic space ensures energy conservation between the initial and final states; if the two far objects have each mass $m$  (as calculated from a $\mathcal{O}(10^{57})$-nucleon neutron star) and the merged object with double the nucleon number (also a preserved quantity) mass $M$, then at least $E=M-2m$ energy is available for emission 
and this can be much larger than in General Relativity, as are the masses of the participating neutron stars.
Indeed, the first released event in the LIGO collaboration has concluded~\cite{LIGO} that the merging objects must be black
holes because the measured total mass seems to be $70m_\odot$ and the mass loss about 3$m_\odot$. Analogous resultshave been recently released by the same collaboration for a second merging event. \cite{LIGO2}.

If gravity is significantly modified at neutron star scale, which certainly remains to be seen, then $f(R)$ theory can potentially accommodate a 3-4 solar mass emission without resort to black-hole merging. Further investigation is very much needed, also taking into account the aforementioned effect of energy emission in the scalar mode and its relative importance with respect to the tensor (gravitational waves) counterpart.

\section{Conclusions}
\label{sec:Conclusions}
%
In this paper, we have investigated the effect of fourth-order $f(R)$ gravity theories in static and spherically symmetric compact neutron star configurations, aiming to settle bounds on the maximum masses  reachable in this kind of theories when realistic equations of state for neutron matter, such as those of~\cite{Gezerlis:2013ipa,Lacour:2009ej,Hebeler:2013nza,Dobado:2011gd,Manuel:2011ed},
are considered.

Our aim was therefore to find  in $f(R)$ theories solutions of equations analogous to that of Tolman-Oppenheimer-Volkov which in General Relativity governs the pressure inside the star through a first-order differential equation. We find that the dynamical system is more entangled in $f(R)$ theories and we are forced to solve higher-order differential equations involving the pressure together with  the metric coefficients and the Ricci scalar. 
This fact renders the solution of the problem much more difficult, even for a fixed $f(R)$ model chosen {\it a priori}. 
Nevertheless we have succeeded, with adequate choice of variables, to reduce the system to six first-order coupled differential equations.

 In the bulk of the article we have also carefully explained that matching between initial conditions at the center of the star where integration begins, and boundary conditions at the border of the star and especially at infinite distance thereof, where the Newtonian potential is to be recovered, requires a numerical shooting-method implementation.

One of our key findings is that other works in the literature have inappropriately
used  the integrated matter-energy of the star's interior as the gravitational mass seen by a far observer (presumably on Earth). This integral in Appendix \ref{App:B}, Eq.~(\ref{massinGR}), is perhaps a convenient label tagging the solutions, but not more: previous treatments naively assumed the Newtonian matching to follow from an outside Schwarzschild-like solution. 

Our Sec.  \ref{sec:mass}  is devoted to a careful explanation of the method implemented which enables us to assign a physical mass to each solution in the asymptotic region where oscillations of the curvature scalar finally damp enough to start retrieving Schwarzschild-like features.

While prior works concluded that astrophysical detection of stars with masses above two solar masses probably constrains extended-gravity $f(R)$ theories with the mass naively defined
as the matter integral over the star, 
we find that our more sophisticated matching yields apparent masses larger than in General Relativity and above such observational values (with numerical size depending on all of the 
specific $f(R)$ theory, its model parameters, and the equation of state under consideration), so that as seen in the bulk of the article, $f(R)$ theories receive no constraint from neutron stars if $\gamma<0$.  This fact can be understood as an emerging effect assignable to the additional $f(R)$ scalar mode which can help to prevent the gravitational collapse and increase the acquirable mass, as well as the Ricci scalar damped oscillations whose energy (outside the star, and often outside its Schwarzschild radius) contributes to the measurable mass as was explained in the bulk of the article.

We report extensive runs of our numerical code determining the viability of such stars in two classes of $f(R)$ models: an inflation of effective-theory motivated quadratic model,  and the popular Hu-Sawicki for later cosmology. 
The treatment of the \emph{quadratic $aR^2$ model}  was partially addressed in previous literature and we contribute an improved labelling of masses, a study with the most updated equations of state, and a full numerical treatment with no perturbative approximation (though we have also sometimes employed perturbation theory to gain insight).
In our investigation we have shown how the bigger the departure from General Relativity (by increasing $a$, the only free parameter of the model), the heavier the stars that can be found for the same radius, easily overcoming the General Relativity Tolman-Oppenheimer-Volkov limit.

The dependence  with the EoS of the Mass-Radius diagrams in figures~(\ref{Figure_Mass_Radius_Diagram}) and (\ref{Figure_Mass_Radius_Diagram2})
 is manifest, encouraging further research in the nuclear aspects to sharpen the statements on the gravitational-theory side. It should be noted how the nuclear equation of state is only extrapolated over a factor $\sim 2$ above nuclear saturation density, whereas General Relativity, if applied, is extrapolated many orders of magnitude in field intensity, in the star's interior, above the current tests based on the exterior metric.

Next, in the \emph{Hu-Sawicki model} we have swiped the parameter space, finding that the relevant one for cosmological applications (small $d$) receives no constraints  in the neutron star context (they are opposite limits of small and large $R$ respectively). 
We have illustrated our analysis thereof with an exponent $n=1$ in Eq.~(\ref{1.3}),
although generalisation for higher exponents is straightforward once the constraints derived from solution stability are imposed. 

We are therefore led to the conclusion that 
after formulating the (relatively involved) system of dynamical equations, the use of realistic neutron fluid EoS, numerical treatment, consistency tests and a correct definition of the mass, compact configurations in the frame of fourth-order $f(R)$ gravity theories present a rich phenomenology that permits large stellar masses, well above what current observations (and General Relativity with the same EoS) would seem to allow.
To decide how large an astronomically discovered mass would signal the need for a modification of GR hinges on further progress in nuclear and particle theory so as to more precisely determine the maximum TOV limit, which cannot be too far above 2.2 solar masses. Significantly higher neutron star masses would make $f(R)$ stars very appealing.

Moreover, we have observed that the neutron star radii (in the sense of the end of the matter distribution, where the pressure drops to zero) are smaller in $f(R)$ theories than in GR. This is due to much of the apparent mass being distributed in the outer gravitational field in $f(R)$ theories, and thus not needing additional matter shells in the star. 
While current measurements of neutron star radii~\cite{Guillot:2014lla} in quiescent X-ray binary systems do point to radii smaller than preferred in General Relativity (9-11 km versus 12-13 km for typical masses),
more accurate measurements are eagerly awaited.

Finally, because the calculated neutron star masses can be much larger than in General Relativity, the energy available for gravitational wave emission as well as the 
total system mass can well exceed what is assumed in GR. Thus, the conclusion of the LIGO collaboration~\cite{LIGO} that the merging objects must be black
holes because the measured total mass seems to be $70m_\odot$ and the mass loss about 3$m_\odot$ is restricted to standard General Relativity.

If gravity is significantly modified at neutron star scale (which certainly remains to be seen) then $f(R)$ theory may accommodate a 3-4 solar mass emission in the merger of neutron stars. This is being investigated.

\begin{acknowledgments}
We thank authors of 
~\cite{Manuel:2011ed} (used in Sec.~\ref{sec:numerics}) and~\cite{Hebeler:2013nza} (used in Secs.~\ref{SectionRsq} and~\ref{SectionHS}) for providing us their numerical data for the corresponding equations of state.
We would like to thank Antonio Dobado, Radouane Ganouji, Antonio L. Maroto and Sergei Odintsov  for the comprehensive reading of the manuscript.
This work has been supported by grants MINECO:FPA2014-53375-C2-1-P, MINECO:FIS2014-52837-P, UCM:910309, Consolider-Ingenio MULTIDARK CSD2009-00064 and CPAN.
Additionally, A.d.l.C.D. acknowledges partial financial support from 
University of Cape Town Launching Grants Programme and 
National Research Foundation grant 99077 2016-2018, Ref. No. CSUR150628121624 and 
 CSIC I-LINK1019.

A.d.l.C.D. also thanks the CANTATA/CA15117 action supported by COST (European Cooperation in Science
and Technology) and the hospitality of the Yukawa Institute for Theoretical Physics (Kyoto, Japan) and the Instituto de F\'isica Te\'orica (IFT UAM-CSIC, Madrid Spain) for its support
via the KA107 action 
for international Mobility during the latest stages of the manuscript. 
F.J.Ll-E. thanks the hospitality of the Institute for Nuclear Theory of the Univ. of Washington, Seattle, with DOE support, as well as the 
Spanish Excellence Network on Hadronic Physics FIS2014-57026-REDT.
\end{acknowledgments}

\appendix
\section{Reduction of the equations of motion and analysis of initial conditions}\label{App:A}
Using the metric provided in Eq.~(\ref{4}), we compute the Ricci tensor, 
and eventually substitute it in the $f(R)$ field equations in Eqs.~(\ref{2}). 
There remain only three independent equations, namely
\begin{eqnarray}\label{5a}
\frac{B'}{4A}\left(\frac{A'}{A}+\frac{B'}{B}\right)-\frac{B''}{2A}-\frac{B'}{rA}=\frac{1}{1+f_{R}}\,[-\kappa\rho B\nonumber\\
+B\left(\frac{A'}{2A^{2}}-\frac{2}{rA}\right)f'_{R}-\frac{B}{A}f''_{R}+\frac{B}{2}(R+f(R))],
\end{eqnarray}
\begin{eqnarray}\label{6a}
\frac{B''}{2B}&+&\frac{B'}{4B}\left(\frac{A'}{A}+\frac{B'}{B}\right)-\frac{A'}{rA}=\frac{1}{1+f_{R}}\,[-\kappa p A\nonumber\\
&+&\left(\frac{B'}{2B}+\frac{2}{r}\right)f'_{R}-\frac{A}{2}(R+f(R))],
\end{eqnarray}
\begin{eqnarray}\label{7a}
&&\frac{1}{A}-1+\frac{r}{2A}\left(-\frac{A'}{A}+\frac{B'}{B}\right)=\frac{1}{1+f_{R}}\,[-\kappa p r^{2}+\frac{r^{2}}{A}f''_{R}\nonumber\\
&&+\left(\frac{B' r^{2}}{2AB}-\frac{A' r^{2}}{2A^{2}}+\frac{r}{A}\right)f'_{R}-\frac{r^{2}}{2}(R+f(R))],\nonumber\\
&&\end{eqnarray}
where the primes denote radial derivatives.
We have used  the perfect fluid $T_{\mu\nu}=(\rho+p)u_{\mu}u_{\nu}-g_{\mu\nu}p$, 
 with $\rho$ and $p$ its density and pressure. The conservation of this tensor provides Eq.
 (\ref{8})
closing the system.  Our philosophy will be to isolate the largest derivative of each independent dynamical quantity to formulate an initial-value system to be solved numerically.
The combination of equations 
$\frac{(\ref{5a})}{2B}+\frac{(\ref{6a})}{2A}+\frac{(\ref{7a})}{r^{2}}$, provides
\begin{eqnarray}\label{9a}
\frac{1}{Ar^{2}}-\frac{A'}{rA^{2}}-\frac{1}{r^{2}}=\frac{1}{1+f_{R}}\,\left[-\frac{\kappa}{2}\left(\rho+3p\right)+\frac{1}{2A}f''_{R}\right.  \nonumber\\
\left.+\left(\frac{3B'}{4AB}-\frac{A'}{4A^{2}}+\frac{1}{rA}\right)f'_{R}-\frac{1}{2}\left(R+f\left(R\right)\right)\right].\ \ \ 
\end{eqnarray}
To isolate $A'$ we need to express $f''_{R}$ in terms of derivatives of smaller order. So a new combination of equations, namely $\frac{3A}{r^{2}}(\ref{7a})+\frac{A}{B}(\ref{5a})-2A(\ref{9a})$, and the use of the definition of the curvature scalar, yields
\begin{eqnarray}\label{10a}
R&=&\frac{B'}{2AB}\left(\frac{A'}{A}+\frac{B'}{B}\right)-\frac{B''}{AB}-\frac{2B'}{rAB}\nonumber\\
&&+\,\,\frac{2A'}{rA^{2}}-\frac{2}{Ar^{2}}+\frac{2}{r^{2}},
\end{eqnarray}
and therefore
\begin{eqnarray}\label{11a}
f''_{R}&=&\left(1+f_{R}\right)\left(\frac{A}{2}R-\frac{A'}{2rA}-\frac{2A}{r^{2}}+\frac{2}{r^{2}}+\frac{3B'}{2rB}\right)
\nonumber\\
&&+\left(\frac{A'}{2A}+\frac{1}{r}\right)f'_{R}\,.
\end{eqnarray}
This last expression is plugged in Eq.~(\ref{9a}) and the result is then rearranged to obtain an equation for $A'$. Once
we introduce the second derivative of $f$ as $f'_{R}=f_{2R}\,R'$, 
the outcome is Eq.~(\ref{5}), the first of our system.

In order to obtain the equation for $R''$, let us return to the trace Eq.~(\ref{3}). Thus by using 
the fact that $f''_{R}=f_{3R}\,R'^{2}+f_{2R}\,R''$ and isolating the term $R''$ therein, Eq.~(\ref{6}) is obtained.
%
%
The last two highest derivatives required to close the system are directly obtained from Eq.~(\ref{6a}) and~(\ref{8}), yielding Eqs.~(\ref{7}) and~(\ref{8}).


At this stage, one notes that $A'$ appears in the right-hand sides of both Eqs.~(\ref{7}) and~(\ref{6}). This function can be obtained in the left-hand side  of Eq.~(\ref{5}). Therefore these three equations require to 
be iterated in our numerical method in the same order as presented in  Sec. \ref{Section2}.

Thus, the set of four Eqs.~(\ref{5})-(\ref{8})
together with the trivial ones for smaller derivatives, form  a system of six first-order differential equations. To start off its numerical integration outwards from the center of the star, we need to specify six initial conditions, namely $\{A, B, B', R, R', p\}$ evaluated at $r=0$. 
The pressure, as a physical quantity, needs to be finite everywhere for the static star to exist, and in particular at $r=0$. Since the system is highly coupled, we need to impose regularity at $r=0$ 
for all equations to avoid any divergence propagating to the pressure. 
Thus proceeding to Eq.~(\ref{7a}) and taking the limit $r\stackrel{}{\rightarrow}0$,
we directly find  $A(0)=1$. Inspecting now Eqs.~(\ref{5a}) and~(\ref{6a}) 
in the same limit, we obtain
\begin{eqnarray}\label{21a}
B'(0)= -A'(0)\,B(0) 
= \frac{2f_{2R}\,R'(0)}{(1+f_{R})}\,B(0)\,,
\end{eqnarray}
due to the pressure expression Eq. (\ref{8}) in order to guarantee a smooth pressure profile at the origin.
In fact, the pressure at the origin $p(0)$ remains a free parameter that characterizes the star providing points in the Mass-Radius diagrams, 
so that we have swiped over its value to generate families of stars.
Thus we conclude that four of the required initial conditions are
$A(0)=1$, $B'(0)=0$, $p(0)=p_{0}$ and $R'(0)=0$; there are only two remaining free initial conditions, namely 
 $B(0)$ and $R(0)$. The latter two are used, with the shooting method, to satisfy appropriate boundary conditions as $r\to\infty$ so the star exterior solution can be matched to the Schwarzschild one at infinity.

$ $\\

\section{Integral form of the gravitational mass}\label{App:B}
In Newtonian theory, the density of matter is defined as the infinitesimal mass element 
divided by the infinitesimal volume, $\rho={\rm d}m/{\rm d}V$, such that the total mass is an integral, that for a purely radial function is
\begin{eqnarray}\label{22b}
M_{Newton}=\int_{0}^{r_{\odot}}\,4\pi r^2 \rho_{matter}(r) {\rm d}r\,, 
\end{eqnarray}
where $r_{\odot}$ stands for the radius of the spherical body and $\rho_{matter}(r)$ for the density of matter, which is a function of the radial coordinate.
In GR, the gravitational mass is obtained solving the field equations and identifying the mass constant by matching the interior and exterior solutions; for the the latter, the metric coefficient $g^{00}$ is then approximated by a Newtonian potential at infinity, with the outcome 
\begin{eqnarray}\label{massinGR}
M_{\rm GR}=\int_{0}^{r_{\odot}}\,4\pi r^2 \rho_{\rm Rel}(r) {\rm d}r\ .
\end{eqnarray}
Now, $\rho_{\rm Rel}(r)=\rho_{\rm matter}(r)+\rho_{\rm energy}(r)$ is the matter-energy density, since both participate as field sources in the relativistic formulation. This is the first and well-known difference respect to the Newtonian gravitational mass. 
But  there is also an additional conceptually important difference: whereas in the Newtonian case, the gravitational mass is the matter density integral over a certain volume, in GR the equivalent integral over matter and energy in Eq.~(\ref{massinGR})
does not play the same role. This is because the product $A(r)B(r)$ inside of star is not equal to one, then $\sqrt{|g|} \neq r^2 \, \sin \theta$, so this factor loses its interpretation as volume element. Consequently the fact that the Schwarzschild mass coincides with Eq.~(\ref{massinGR}) must be regarded as a pure coincidence. In general, Eq.~(\ref{massinGR}) must be then regarded just as a parameter characterizing a family of solutions in GR, with no claim of it being a physical mass. An instance where this difference matters is a binary system. Two compact stars, both with the mass of Eq.~(\ref{massinGR}), have a very different behavior than a test mass in a Schwarzschild metric as controlled by Eq. (\ref{23}).


\end{document}